\documentclass[smallextended]{svjour3} 
\usepackage{amsmath}
\usepackage{latexsym}
\usepackage{amssymb}
\usepackage{graphics,epstopdf}
\usepackage{graphicx}
\usepackage[colorlinks=true, citecolor=blue, urlcolor=blue ]{hyperref}
\usepackage{float}
\usepackage{graphicx}
\usepackage{amsfonts}
\usepackage{subcaption}
\usepackage{tabularx, booktabs}

\usepackage{mathtools}
\usepackage{cite}

\newcommand{\be}{\begin{equation}}
\newcommand{\ee}{\end{equation}}
\newcommand{\ba}{\begin{eqnarray}}
\newcommand{\ea}{\end{eqnarray}}

\allowdisplaybreaks
\usepackage[dvipsnames]{xcolor}
\newtheorem{sce}{Scenario}

\begin{document}
	\title{Resource theoretic efficacy of the single copy of a two-qubit entangled state in a sequential network}
	
	\author{Arun Kumar Das \and Debarshi Das\footnote{Corresponding Author}   \and Shiladitya Mal   \and Dipankar Home \and
		A. S. Majumdar
	}

\institute{Arun Kumar Das \at S. N. Bose National Centre for Basic Sciences, Block JD, Sector III, Salt Lake, Kolkata 700 106, India \\
	\email{akdkumar1994@gmail.com}  
	\and
	Debarshi Das \at S. N. Bose National Centre for Basic Sciences, Block JD, Sector III, Salt Lake, Kolkata 700 106, India 
	\at Department of Physics and Astronomy, University College London, Gower Street, WC1E 6BT London, England, United Kingdom \\
	\email{dasdebarshi90@gmail.com}  
	\and 
	Shiladitya Mal \at Department of Physics and Center for Quantum Frontiers of Research and Technology (QFort),
	National Cheng Kung University, Tainan 701, Taiwan
\at Physics Division, National Center for Theoretical Sciences, Taipei 10617, Taiwan \\
	\email{shiladitya.27@gmail.com}      
	\and 
	Dipankar Home \at Centre for Astroparticle Physics and Space Science, Bose Institute, Block EN, Sector V, Salt Lake, Kolkata 700 091, India \\
	\email{quantumhome80@gmail.com}        
	\and 
	A. S. Majumdar \at S. N. Bose National Centre for Basic Sciences, Block JD, Sector III, Salt Lake, Kolkata 700 106, India\\
	\email{archan@bose.res.in}
}

	\maketitle

	\begin{abstract}
		How best one can recycle a given quantum resource, mitigating the various
		difficulties involved in its preparation and preservation, is of
		considerable importance for ensuring efficient applications in quantum
		technology.   Here we demonstrate quantitatively the resource theoretic 
		advantage of reusing a single copy of a two-qubit entangled state towards
		information processing. 
		 To this end, we consider a scenario
		of sequential entanglement detection of a given two-qubit state by multiple independent observers on each of the two spatially separated wings. In particular,  we
		consider equal numbers of sequential observers on the two wings. We first determine the upper bound on the number of observers who can detect entanglement employing  suitable entanglement witness operators. In terms of the parameters
		characterizing the entanglement consumed and the robustness of measurements,   we then compare  the above scenario  with the corresponding scenario involving multiple pairs of
  entangled qubits shared among the two wings. This reveals a clear resource theoretic advantage of
		recycling a single copy of a two-qubit entangled state in the sequential network. 
	\end{abstract}

	\section{Introduction}
	One of the most counterintuitive features of quantum mechanics is quantum entanglement \cite{ent1,ent2,ent3}, which leads to nonclassical phenomena like Bell nonlocality \cite{belln1,belln2} and Einstein-Podolsky-Rosen steering \cite{eprs1,eprs2,eprs3}. Apart from the foundational significance of probing incompatibility between quantum mechanical predictions and the local realist descriptions of nature, quantum entanglement serves as a resource for various information processing and communication tasks. To name a few, some such
	well-established tasks include quantum teleportation \cite{teleport}, quantum dense coding \cite{dense_code}, quantum key distribution \cite{qkd}, certification of genuine randomness \cite{rand}, and quantum random access codes \cite{rac1}. 
	
	 In a real laboratory set-up, preparation of any quantum resource always faces different types of complications \cite{prep1} and such difficulties are quantified by the  ``preparation cost" associated with the preparation dynamics \cite{prep2}. Moreover, it is an extremely difficult task to
	prepare quantum resources having a high degree of isolation from
	environmental interactions  \cite{env1,env4}. In fact, quantum correlations in independent environments have been shown to decay asymptotically or even to disappear at a finite time under the action of noise \cite{env2,env3,env5}. Therefore, the efficient use of quantum resources is one of the primary challenges in the backdrop of current endeavour of building quantum technology.
	To this end, the possibility of recycling the same resource several times is of great advantage.
	A particular network scenario suitable for this purpose comprises a
	single copy of a bipartite entangled state with multiple pairs of
	independent sequential observers in the two spatially separated wings,
	where each of these observers performs unsharp measurement and delivers the
	accessed particle to the next observer \cite{silva_prl}. 
	
	Specifically, in~\cite{silva_prl}, the authors  showed that at most two independent observers
	at one wing can violate the Bell-CHSH (Clauser-Horne-Shimony-Holt) inequality \cite{CHSH} with an observer on the other wing by sharing a pair of entangled
	spin-$1/2$ particles.  Note that all except the last one in a sequence of 
	multiple observers cannot perform sharp or projective measurements, since it
	is desired that some amount of entanglement must survive in the post-measured state in order
	to be utilized by the subsequent observer.  It was shown in~\cite{mathematics_mal}, that the upper bound of two observers on one wing who
	can share nonlocality of a two-qubit state in the scenario of unbiased measurement settings, is based on the optimality of the
	unsharp measurement framework \cite{um1,unsharp_measurement}  with respect to the trade-off between information gain and disturbance  in a quantum measurement \cite{igd1,igd2}. It is thus important to employ such measurements in order to obtain optimal performance in the above sequential network scenario.
	
	The issue of sequential detection of different quantum correlations by multiple observers has been investigated both theoretically and experimentally in several subsequent works using the unsharp measurement formalism. Such works
	include, for example, steering a single system multiple times \cite{sasmal_steering,shenoy,choi_steering,shashank_steering,steernew21,estwo,estwo2}, exploring Bell-type nonlocality in various settings \cite{npj_2018,Qu2017,debarshi_facets,rennew,Saha19,foletto_entanglement,expnew,zhang21,cglmp,prl_brown,hallnew}, witnessing entanglement of bipartite and tripartite
	states \cite{bera_entanglement,ananda_entanglement,srivastava_entanglement}, sharing of nonlocal advantage of quantum coherence \cite{coherence_sounak},
	and quantum contextuality \cite{akpan}. Applications of sequential detection of quantum correlations in different information processing tasks have also been reported,  e.g., in the context of randomness certification \cite{rantt}, dimension witness \cite{appln1tt}, quantum random access codes \cite{ractt,appln4tt,appln5tt},  quantum teleportation  \cite{sroytt}, remote state preparation \cite{rsp_shounak},  distinguishing
quantum predictions from classical simulations with finite memory \cite{cctt}.  	Recently, the technique of choosing different sharpness parameters for the different measurement settings of each observer has been proposed for
	obtaining the possibility of unbounded number of observers sharing Bell-nonlocality \cite{prl_brown,bntwo,cabello}, and this type of result has also been probed towards random access code generation \cite{rac_debarshi}.

	The above works have stimulated wide interest in recycling various types of quantum
	correlations for their use in multi-observer networks. A natural question
	emerges in this context as to if any quantitative advantage can be gained from the
	resource theoretic perspective by the reuse of correlations in a single
	copy of a quantum state.  In the present work we answer this question in
	the affirmative.  In particular, we consider a single
	copy of a bipartite two-qubit entangled state that is shared between  multiple observers on both the wings, who sequentially and independently perform measurements on the
	state. We first determine the essential figure of merit in this scenario, that is given by the maximum number of observers who can successfully detect entanglement contained in the bipartite two-qubit state. We next focus towards addressing
	resource theoretic comparison
	of performance of sequential network schemes based on single copy entangled state, with that of schemes based on multi-copy entangled states shared by
	the same number of observers. In order to compare the sequential and the non-sequential scenarios  we utilize the detectability (or visibility)  in terms
	of the expectation value of the entanglement witness operator \cite{ewo,oew1}, 
	as well as the  information extraction capability of the involved measurements, defined quantitatively  via the robustness of measurement \cite{rom_prl}.   
	In terms of the above two parameters, we show that  the sequential measurement 
	protocol provides advantages in witnessing entanglement by multiple pair of observers in terms of the resources consumed.
	

	We arrange the rest of the paper in the following way. In Sec. \ref{sec2}, we 
	provide a brief overview of some basic tools employed in our analysis, such as  entanglement witness operators, unsharp measurement and robustness of measurements. Next, in Sec.  \ref{sec3}, we provide details of our symmetric  network scenario. The main results regarding the bounds on the
	number of observers, and the resource theoretic comparisons  are discussed in Sec. \ref{sec4}.  In Sec. \ref{teleportation}  we illustrate the resource theoretic advantage of the 
	sequential scenario through an example of quantum teleportation  using witness operators for detecting useful entangled states for quantum teleportation. Sec. \ref{sec5} contains
	an analysis of the asymmetric extension of the above scenario. Finally, in Sec. \ref{sec6} we  summarize our results with some concluding discussions.

	\section{Basic tools} \label{sec2}
	
	In this section, we present the basic ideas of entanglement witness operators, unsharp measurements and robustness of measurements. These concepts will be used later for presenting the main results of this paper. 
	
	\subsection{Entanglement Witness Operators}
	
	A Hermitian operator $W$ is called an entanglement witness operator if there exists at least one entangled state $\rho_e \notin \mathcal{S}$ such that $\text{Tr}(W\rho_e )<0$ and $\text{Tr}(W\rho) \ge 0$ for all $\rho \in \mathcal{S}$ with $\mathcal{S}$ being the set of all separable states  \cite{ewo}. 
	One can find out an entanglement witness operator  for each entangled state. However, finding out the optimal entanglement witness operator for a given entangled state is not always  easy \cite{oew1}. For detecting entanglement, one is usually interested in decomposing an entanglement witness operator in terms of local quantum measurements. This enables performing the detection process using local quantum measurements \cite{ewo}.
	
	Consider the state $|\psi^{+}\rangle = \frac{1}{\sqrt{2}} (|01\rangle + |10\rangle)$. The optimal entanglement witness operator for this state is given by \cite{ewo},
	\begin{equation}\label{eq1}
	W=\frac{1}{4} \Big( \mathbb{I} \otimes \mathbb{I} +\sigma_{z} \otimes \sigma_{z} -\sigma_{x} \otimes \sigma_{x} - \sigma_{y} \otimes \sigma_{y} \Big).
	\end{equation}	
	The advantage of this entanglement witness operator is that it can be implemented in the laboratory by performing a three correlated local quantum measurements in the bases associated with the Pauli operators $\{\sigma_x, \sigma_y, \sigma_z\}$. 
	
	If in the preparation process of the state $|\psi^{+}\rangle$
	some random noise acts, then the resultant state may turn out to be the Werner state of the form,
	\begin{equation}	\label{werner_state}
	\rho = p |\psi^{+}\rangle \langle \psi^{+}| + (1-p) \frac{\mathbb{I}}{2} \otimes \frac{\mathbb{I}}{2},
\end{equation}
	where $0 < p \leq 1$; $\frac{\mathbb{I}}{2} \otimes \frac{\mathbb{I}}{2}$ denotes white noise and $(1-p)$ is the strength of the noisy process. The entanglement witness operator $W$ given by Eq.(\ref{eq1}) remains optimal for the state $\rho$ as well \cite{ewo}. 
	
	\subsection{Optimality of Unsharp Measurements}
	
	A fundamental feature of quantum theory is that no information about a
	system can be obtained without perturbing its state \cite{heisenberg}.  Projective measurements, also known as strong measurements, are the most informative at the cost of maximally disturbing the initial state. On the other hand, weak measurements are characterised by broad pointer states, and are less informative, but affect 
	lesser the initial state \cite{silva_prl}, thereby reflecting a nontrivial trade off between information gain and disturbance  \cite{igd1,igd2}. In our sequential networks, it is important to employ such weak measurements which optimize the information gain-disturbance trade-off to achieve best performance, rather than some random choice of measurement. In \cite{mathematics_mal}, it was shown that unsharp measurement which is an one-parameter Positive Operator-Valued Measure (POVM) satisfies the optimality criteria. 
	
	Generalized quantum measurement or POVM \cite{um1,unsharp_measurement} is defined by a set of positive operators that add to identity, i.e., $E \equiv \lbrace E_{i}\vert\sum E_{i}=\mathbb{I},0 \leq E_i\leq \mathbb{I}\rbrace$.  
	Consider the dichotomic observable $\vec{\sigma} \cdot \hat{n}$, which is the spin component observable for qubits along the direction $\hat{n}$. Here $\vec{\sigma} = (\sigma_x, \sigma_y, \sigma_z)$ is a vector composed of three Pauli operators and $\hat{n}$ is a unit vector in $\mathbb{R}^3$. 
	Given the observable $\vec{\sigma} \cdot \hat{n}$, one can define the dichotomic unsharp observable  $E^{\lambda}_{\hat{n}} = E^{\lambda}_{+|\hat{n}} - E^{\lambda}_{-|\hat{n}}$ \cite{uno2} associated with the sharpness parameter $\lambda \in (0, 1]$, where 
	\begin{equation}
	E^{\lambda}_{\pm|\hat{n}} = \lambda P_{\pm|\hat{n}} + \frac{1-\lambda}{2}\,\, \mathbb{I}
	\label{unsharpm}
	\end{equation}
	are the effect operators that satisfy $E^{\lambda}_{+|\hat{n}} + E^{\lambda}_{-|\hat{n}} = \mathbb{I}, \, 0 \leq E^{\lambda}_{\pm|\hat{n}} \leq \mathbb{I}$. Here, $P_{\pm|\hat{n}}$ are the projectors given by, $P_{\pm|\hat{n}} = (\mathbb{I} \pm \vec{\sigma} \cdot \hat{n})/2$.

	The probabilities of getting the outcomes $+1$ and $-1$, when the above unsharp measurement is performed on the state $\rho$, are given by $\text{Tr}[\rho E^\lambda_{+|\hat{n}}]$ and $\text{Tr}[\rho E^\lambda_{-|\hat{n}}]$, respectively. 
	The expectation value of $E^{\lambda}_{\hat{n}}$ for a given $\rho$ is defined as,
	\begin{eqnarray}
	\langle E^{\lambda}_{\hat{n}} \rangle = \text{Tr}[\rho E^\lambda_{+|\hat{n}}] - \text{Tr}[\rho E^\lambda_{-|\hat{n}}]= \lambda \langle \vec{\sigma} \cdot \hat{n} \rangle,
	\end{eqnarray}
	where $\langle \vec{\sigma} \cdot \hat{n} \rangle$ $=$ $\text{Tr}[\rho (P_{+|\hat{n}} - P_{-|\hat{n}})]$  denotes the expectation value of the observable $\vec{\sigma} \cdot \hat{n}$ under projective measurement. 
	The post-measurement state can be determined using the generalized von Neumann-L\"{u}ders transformation rule \cite{um1,unsharp_measurement} as follows,
	\begin{equation}
	\rho\rightarrow\frac{\sqrt{E^{\lambda}_{\pm|\hat{n}}} \, \rho \, \sqrt{E^{\lambda}_{\pm|\hat{n}}}}{\text{Tr}(\rho E^{\lambda}_{\pm|\hat{n}})}.
	\end{equation}
	In the weak measurement formalism with broad pointer states, the optimal information gain-disturbance trade-off is characterised by a condition involving two parameters called quality factor (F) and precision (G). An optimal pointer satisfies $F^2+G^2=1$, which implies maximal information gain for a given amount of disturbance \cite{silva_prl}. In the unsharp measurement formalism described above, it turns out that $G=\lambda, F=\sqrt{1-\lambda^2}$ \cite{mathematics_mal}, thereby satisfying the optimality condition.
	
	\subsection{Robustness of Measurements}
	
	The concept of ``Robustness of Measurements'' (RoM) has been formulated 
	recently to quantify the informativeness of a measurement \cite{rom_prl}.   Given a
	particular measurement $E$, RoM of $E$, denoted by $R(E)$, quantifies to what extent
	$E$ is a resourceful measurement. 
	When a measurement	 returns an arbitrary outcome $i$ with probability $q(i)$
	independent of the quantum state measured, the measurement is called a trivial measurement. Such a
	measurement has POVM elements $E_i$ with $E_i= q(i)\mathbb{I}$ for all $i$.  It is evident that trivial measurements are not informative or resourceful at all. 
	
	RoM	is defined as the minimal amount of noise that needs to be added to the measurement such that the measurement  becomes a trivial one.  Suppose, instead of always performing the measurement  $E=\{E_{i}\}$, one performs a different measurement  $F=\{F_{i}\}$ sometimes. The informativeness of the measurement $E$ can be captured by the minimal probability of this other measurement $F$  that makes the overall measurement trivial. Hence, the RoM can be formally defined as \cite{rom_prl},
	\begin{align}
	&R(E) = \min_{F,q} r, \nonumber \\	
&\text{such that} \nonumber \\
    &\frac{E_{i}+ r F_{i}}{1+r}=q(i) \, \mathbb{I}  \hspace{0.4cm} \forall \, \, i, \nonumber \\
	&  F_i \geq 0 \, \, \forall \, i \, \, \, \text{and} \, \, \, \sum_i F_i = \mathbb{I}.
	\end{align}
	Here, the minimization is taken over all noise measurements  $F=\{F_{i}\}$ and all probability distributions $q=\{q(i)\}$.     
	
RoM can also be written as \cite{rom_prl},
	\begin{equation}
	R(E)=\sum_{i}||E_{i}||_{\infty} - 1,
	\end{equation}
	where $ ||E_i||_{\infty}$ is the operator norm of $E_i$. Note that $ ||E_i||_{\infty}$ is equal to the maximum eigen value of $\sqrt{E_i^{\dagger} E_i}$ \cite{hayashi_book}.  
	Hence, for the unsharp measurement $E^{\lambda}_{\hat{n}} \equiv \{E^{\lambda}_{+|\hat{n}}, E^{\lambda}_{-|\hat{n}}\}$ defined in Eq.(\ref{unsharpm}), we have  
	\begin{equation}
	R(E^{\lambda}_{\hat{n}}) = \lambda.
	\end{equation}
	Using the resource theory of measurement informativeness \cite{rom_prl}, any trivial measurement can be considered as a free measurement
	and any measurement which is not trivial is a
	resourceful measurement. Hence, RoM characterises the amount of resource in a measurement, i.e., its information extraction capacity.

	\section{Setting up the scenario}\label{sec3} 
	
\begin{figure}
		\centering
		\includegraphics[width=200px,height=130px]{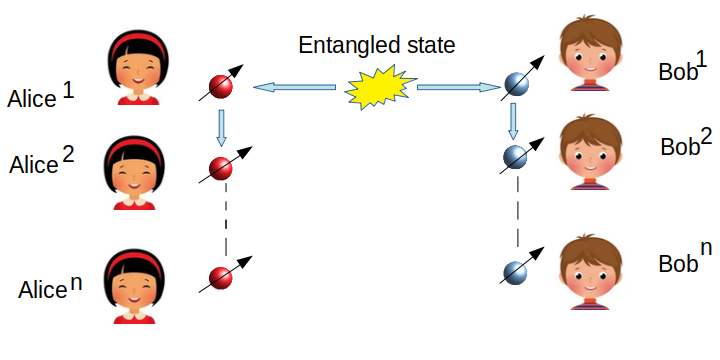}
		\caption{An entangled state $\rho_i$ of two spin-$\frac{1}{2}$ particles is initially shared between Alice$^1$ and Bob$^1$. They perform measurements on the particles in their possession to detect the entanglement, and after the measurements Alice$^1$ sends her particle to Alice$^2$ and Bob$^1$ sends his particle to Bob$^2$. Next, Alice$^2$ and Bob$^2$ perform measurements to detect entanglement of the shared state, and then sends the particles to the next pair, i.e., Alice$^3$-Bob$^3$ and the process continues. The process is terminated when a sequential pair, say, Alice$^m$-Bob$^m$ is unable to detect entanglement of the shared state.}\label{fig1}
	\end{figure}

	In the present study, we will consider the particular scenario as described below (see Fig. \ref{fig1}).
	
	\begin{sce}
		We consider $n$ number of sequential Alices (Alice$^1$, Alice$^2$, Alice$^3$, $\cdots$, Alice$^n$) and $n$ number of sequential Bobs (Bob$^1$, Bob$^2$, Bob$^3$, $\cdots$, Bob$^n$), where $n$ is a-priori arbitrarily large. At first, the pair Alice$^1$-Bob$^1$ detects the entanglement of the initial state $\rho_i$ of two spin-$\frac{1}{2}$ particles using entanglement witness operator. Alice$^1$ then passes her particle to Alice$^2$ and Bob$^1$ passes his particle to Bob$^2$. Alice$^2$-Bob$^2$ then detects entanglement of the two-qubit state received after the measurements by Alice$^1$ and Bob$^1$. Consequently, Alice$^2$ and Bob$^2$ pass their particles to Alice$^3$ and Bob$^3$ respectively, and so on. The process is terminated when  Alice$^m$-Bob$^m$ with $m \leq n$  is unable to detect entanglement.  \nonumber
		\label{sce1}
	\end{sce}

	In  the above scenario  we consider the following  assumptions:
	
	1) Each Alice (Bob) performs measurements independent of the measurement settings and outcomes of the previous Alices (Bobs).
	
	2)  All possible measurement settings of each Alice (Bob) are equally probable. 
	
	3) Each Alice (Bob) employs the same
	value of the sharpness parameter for all of her (his) measurement settings.
	
	In the case of sharp projective measurement, one obtains the maximum amount of information at the cost of maximum disturbance to the state. In the scenarios considered by us, Alice$^i$ (Bob$^i$) passes on the respective particle to Alice$^{i+1}$ (Bob$^{i+1}$) after performing suitable measurement. In this case, Alice$^i$ (Bob$^i$) needs to perform measurement for detecting entanglement by disturbing the state minimally such that some entanglement remains in the post-measurement state to be detected by Alice$^{i+1}$ (Bob$^{i+1}$). This can be achieved in the
	unsharp measurement formalism as the disturbance is minimized for any fixed amount of information gain in this formalism for qubits \cite{mathematics_mal,sasmal_steering}.

	\subsection{Modified entanglement witness operator in unsharp measurement formalism} \label{sec3a}
	
	Note that the entanglement witness operator (\ref{eq1}) can be implemented in the laboratory by performing projective quantum measurements. Since, in our scenarios,  all Alices (Bobs), except the last Alice (Bob) in a sequence,  perform unsharp measurements, the entanglement witness operator (\ref{eq1}) needs to be modified accordingly.
	In order to modify the entanglement witness operator (\ref{eq1}), we  follow the process described in \cite{bera_entanglement,ananda_entanglement}.
	
	Suppose, Alice$^i$ and Bob$^j$ perform unsharp measurement of spin component observables $\vec{\sigma} \cdot \hat{n}_i$ and $\vec{\sigma} \cdot \hat{m}_j$ respectively. The sharpness parameters associated with the measurements by Alice$^i$ and Bob$^j$ are denoted by $\xi_{i}$ and $\lambda_j$ respectively with $\xi_{i}, \lambda_j$ $\in$ $(0,1]$. We will follow this notation throughout the paper.
	
	The joint probability of obtaining the outcomes $a_i$, $b_j$ (with $a_i, b_j \in \{+1, -1\}$), when Alice$^i$ and Bob$^j$ perform the above unsharp measurements, can be evaluated using the expression,
	\begin{equation}
	\mbox{Tr}\Big[\rho \Big(E^{\xi_i}_{a_i|\hat{n}_i}  \otimes E^{\lambda_j}_{b_j|\hat{m}_j} \Big)\Big], 
	\nonumber
	\end{equation}  
	where $\rho$ is the state shared by Alice$^i$ and Bob$^j$; and the expressions of $E^{\xi_i}_{a_i|\hat{n}_i}$  and $E^{\lambda_j}_{b_j|\hat{m}_j}$ are defined following Eq.(\ref{unsharpm}).
	The expectation value of the above joint measurement in the state $\rho$ is given by,
	\begin{align}
	\langle E^{\xi_i}_{\hat{n}_i} \otimes E^{\lambda_j}_{\hat{m}_j} \rangle 
	& = \mbox{Tr}\Big[ \Big\{ \Big(E^{\xi_i}_{+|\hat{n}_i}-E^{\xi_i}_{-|\hat{n}_i} \Big) \otimes \Big(E^{\lambda_j}_{+|\hat{m}_j} - E^{\lambda_j}_{-|\hat{m}_j} \Big)  \Big\}  \rho \Big] \nonumber \\
	& = \xi_i \, \lambda_j \, \mbox{Tr}\Big[ \Big\{ \Big(P_{+|\hat{n}_i}-P_{-|\hat{n}_i} \Big) \otimes \Big(P_{+|\hat{m}_j} - P_{-|\hat{m}_j} \Big)  \Big\}  \rho \Big] \nonumber \\
	& = \xi_i \, \lambda_j \, \langle \vec{\sigma} \cdot \hat{n}_i \otimes \vec{\sigma} \cdot \hat{m}_j \rangle,
	\end{align}
	where $\langle \vec{\sigma} \cdot \hat{n}_i \otimes \vec{\sigma} \cdot \hat{m}_j \rangle$ is the expectation value under projective measurements. We can 
	hence use the substitution $\langle \vec{\sigma} \cdot \hat{n}_i \otimes \vec{\sigma} \cdot \hat{m}_j \rangle \rightarrow \xi_i \, \lambda_j \, \langle \vec{\sigma} \cdot \hat{n}_i \otimes \vec{\sigma} \cdot \hat{m}_j \rangle$ in order to obtain the modified entanglement witness operator for the case of unsharp measurements. 
	For any $\xi_{i}, \lambda_j$ $\in$ $(0,1]$, the modified entanglement witness operator for the state $|\psi^{+}\rangle = \frac{1}{\sqrt{2}} (|01\rangle + |10\rangle)$ has the following form,
	\begin{align}
	W^{(\xi_i , \lambda_j)} 
	&=  \frac{1}{4} \Big( \mathbb{I} \otimes \mathbb{I} + \xi_i \sigma_{z} \otimes \lambda_j \sigma_{z} - \xi_i \sigma_{x} \otimes \lambda_j \sigma_{x}  - \xi_i \sigma_{y} \otimes \lambda_j \sigma_{y} \Big) \nonumber \\
	&=  \frac{1}{4} \Big[ \mathbb{I} \otimes \mathbb{I} + \xi_i \lambda_j \Big( \sigma_{z} \otimes  \sigma_{z} -  \sigma_{x} \otimes  \sigma_{x}  -  \sigma_{y} \otimes   \sigma_{y} \Big) \Big]. 
	\label{unwo}
	\end{align}	
	
	Now,  for any separable state $\rho_s \in \mathcal{S}$, we have 
\begin{align}
	\text{Tr} \Big(W^{(\xi_i , \lambda_j)} \rho_s \Big) &= \text{Tr} \Big[ \big(\xi_i \, \lambda_j \, W + \frac{1}{4}(1- \xi_i \, \lambda_j )   \mathbb{I} \otimes \mathbb{I}  \big) \rho_s \Big] \nonumber \\
	& = \xi_i \, \lambda_j \text{Tr}\Big(W \rho_s \Big) + \frac{1}{4} (1- \xi_i \, \lambda_j ). 
	\end{align}
This implies that $  \text{Tr} \Big(W^{(\xi_i , \lambda_j)}\rho_s\Big) \ge 0$ for all $\rho_{s} \in S $ as  $0 < \xi_i,\lambda_i \le 1$. Hence, $W^{(\xi_i , \lambda_j)}$ is a valid entanglement witness operator.

	\section{Witnessing entanglement by sequential observers}\label{sec4} 	
	
	We now focus on the task of entanglement detection in our sequential network
	scenario. We take three different types of the initially shared states $\rho_i$.
	
	\subsection{Initially shared maximally entangled two-qubit state }\label{sec4a}
	
	Let us consider that Alice$^1$ and Bob$^1$ initially share the Bell state given by, $|\psi^{+}\rangle = \frac{1}{\sqrt{2}} (|01\rangle + |10\rangle)$. Different pairs of Alice and Bob (i.e., Alice$^1$-Bob$^1$, Alice$^2$-Bob$^2$, Alice$^3$-Bob$^3$, $\cdots$, Alice$^n$-Bob$^n$) try to detect entanglement sequentially.	At first, we will address the following question: how many such pairs of Alice and Bob can sequentially detect entanglement. 
	
	Let Alice$^i$ and Bob$^i$ ($i \in \{1, 2, \cdots, n\}$) perform unsharp measurements with sharpness parameters $\xi_i$ and $\lambda_i$ respectively. 
	The pair Alice$^1$-Bob$^1$ can detect entanglement if the following condition is satisfied:	
	\begin{equation}
	\text{Tr} \Big[W^{(\xi_{1},\lambda_{1})} |\psi^{+} \rangle \langle \psi^{+}| \Big]<0, 
	\label{cond1}
	\end{equation}
	where the operator $W^{(\xi_{1},\lambda_{1})}$ is given by Eq.(\ref{unwo}). After simplification, we get the following condition from (\ref{cond1}),
	\begin{equation}
	\xi_{1} \, \lambda_{1}>\frac{1}{3}
	\label{con1}
	\end{equation}
	
	Next, let us find out the post measurement state received by the pair Alice$^2$-Bob$^2$ from Alice$^1$-Bob$^1$. As Alice$^2$ (Bob$^2$) acts independent of the measurement setting and outcome of Alice$^1$ (Bob$^1$) in each experimental run, we take average over the  measurement settings and outcomes by Alice$^1$ and Bob$^1$. Hence, the state received, on average, by  Alice$^2$-Bob$^2$ from Alice$^1$-Bob$^1$ is given by,
		\begin{align}
		\rho_{A_{2}B_{2}} =\frac{1}{9}\sum_{n_1, m_1,a_1,b_1} \Big(\sqrt{E^{\xi_{1}}_{a_1|\hat{n}_1}} \otimes \sqrt{E^{\lambda_{1}}_{b_1|\hat{m}_1}} \Big) \, |\psi^{+} \rangle \langle \psi^{+}| \,  \Big(\sqrt{E^{\xi_{1}}_{a_1|\hat{n}_1}} \otimes \sqrt{E^{\lambda_{1}}_{b_1|\hat{m}_1}} \Big) ,
		\label{postst1}
		\end{align}
	with	$\hat{n}_1, \hat{m}_1 \in \{ \hat{x}, \hat{y}, \hat{z}\}$ and $a_1,b_1 \in \{+1,-1\}$. Here, we have used the fact that each of Alice$^1$ and Bob$^1$ performs any of the three local unsharp  measurements associated with the observables $\sigma_x$, $\sigma_y$, $\sigma_z$ in each experimental run in order to implement the entanglement witness operator (\ref{unwo}). We have also used here the assumption that all possible measurement settings of Alice$^1$ and that of Bob$^1$ are equally probable. After simplification, we get from Eq.(\ref{postst1}),
	\begin{align}
	\rho_{A_{2}B_{2}} &= p |\psi^{+}\rangle \langle \psi^{+}| + (1-p) \frac{\mathbb{I}}{2} \otimes \frac{\mathbb{I}}{2} \nonumber \\
	& \text{with} \, \, \, p = \frac{1}{9} \Big(1 + 2 \sqrt{1 - \xi_1^2} \Big) \Big(1 + 2 \sqrt{1 - \lambda_1^2} \Big).
	\label{statea2b2}
	\end{align}
	
	Since, the state (\ref{statea2b2}) has the form given by Eq.(\ref{werner_state}), the Alice$^2$-Bob$^2$ pair again uses the same entanglement witness operator given by Eq.(\ref{unwo}) to detect entanglement. Hence, Alice$^2$-Bob$^2$ can detect entanglement if the following condition is satisfied, 
	\begin{equation}
	\text{Tr} \Big[W^{(\xi_{2},\lambda_{2})} \, \rho_{A_{2}B_{2}} \Big]<0, 
	\label{cond2}
	\end{equation}
	which implies the condition,	
	\begin{equation}
	\xi_{2} \, \lambda_{2} > \frac{3}{\Big(1+2\sqrt{1-\xi_{1}^{2}} \Big) \Big(1+2\sqrt{1-\lambda_{1}^{2}} \Big)}.
	\label{con2}
	\end{equation}
	
	Proceeding in a similar way, it can be shown that the state $\rho_{A_{3}B_{3}}$ received, on average, by  Alice$^3$-Bob$^3$ from Alice$^2$-Bob$^2$ has the similar form  of Werner state (\ref{werner_state}) and the pair	Alice$^3$-Bob$^3$ can detect entanglement if
	\begin{equation}
	\text{Tr} \Big[W^{(\xi_{3},\lambda_{3})} \, \rho_{A_{3}B_{3}} \Big]<0, 
	\label{cond3}
	\end{equation}
	i.e., when
		\begin{equation}
		\xi_{3} \, \lambda_{3} >\frac{27}{\prod\limits_{i=1}^{2} \left[\Big(1+2\sqrt{1-\xi_{i}^{2}} \Big) \Big(1+2\sqrt{1-\lambda_{i}^{2}} \Big) \right]}.
		\label{con3}
		\end{equation} 
		
		Repeating the above steps, it can be shown that 	Alice$^4$-Bob$^4$ can detect entanglement if	the following condition is satisfied,
	\begin{equation}
	\xi_{4} \, \lambda_{4} > \frac{243}{ \prod\limits_{i=1}^{3} \left[ \Big(1+2\sqrt{1-\xi_{i}^{2}} \Big) \Big(1+2\sqrt{1-\lambda_{i}^{2}} \Big) \right]}.
		\label{con4}
		\end{equation} 
		
	Similar conditions can be found out that ensure entanglement detection by the other pairs, i.e., Alice$^n$-Bob$^n$.
	
	Now, our purpose is to investigate what is the maximum number of the sequential pairs succeed in witnessing the entanglement of the shared two-qubit state. Combining Eqs.(\ref{con1}), (\ref{con2}), (\ref{con3}) and (\ref{con4}) and performing some analytical calculations (see Appendix \ref{a1} for details), we get that Alice$^1$-Bob$^1$, Alice$^2$-Bob$^2$ and Alice$^3$-Bob$^3$ can  detect entanglement if the following conditions are satisfied simultaneously,
	\begin{align}
	&\xi_{1} = \lambda_{1} = 0.58 + \delta_1  \hspace{0.25cm} \text{with} \hspace{0.25cm} 0 \leq \delta_1 <<1, \label{sh1} \\
	&\xi_2 = \lambda_2 = 0.66 + \delta_2  \hspace{0.25cm} \text{with} \hspace{0.25cm} 0 \leq \delta_2 <<1, \label{sh2} \\
	& \xi_{3} = \lambda_{3} = 0.79 + \delta_3  \hspace{0.25cm} \text{with} \hspace{0.25cm} 0 \leq \delta_3 <<1. \label{sh3}
	\end{align} 
	Here the numerical digits appearing  in the  the above conditions are rounded to two decimal places.

	If Alice$^1$-Bob$^1$, Alice$^2$-Bob$^2$ and Alice$^3$-Bob$^3$ perform measurements with sharpness  parameters satisfying Eqs.(\ref{sh1}) (\ref{sh2}), (\ref{sh3}) , then it can be shown that Alice$^4$-Bob$^4$ cannot witness the entanglement even if they perform projective measurements, i.e., with $ \xi_{4}=\lambda_{4}=1 $ (see Appendix \ref{a1} for details). Thus, at most three sequential pairs of Alice and Bob can witness the entanglement in this case.

	\subsection{Initially shared two-qubit Werner state}
	
	Let Alice$^1$ and Bob$^1$ initially share the two-qubit Werner state mentioned in Eq.(\ref{werner_state}). For this state, the optimal entanglement witness operator remains the same as before, i.e., it is $W$ given by Eq.(\ref{eq1}) \cite{ewo}. 
	
	In this case  at most three sequential pairs of Alice and Bob (for example, Alice$^1$-Bob$^1$, Alice$^2$-Bob$^2$ and Alice$^3$-Bob$^3$) can witness the entanglement using the entanglement witness operator given by Eq.(\ref{unwo}). Furthermore, we get the following results:
	
	(1) When $ 0.80 < p \le 1 $,  each of the pairs Alice$^1$-Bob$^1$, Alice$^2$-Bob$^2$ and Alice$^3$-Bob$^3$ can detect the entanglement.  Other pairs Alice$^i$-Bob$^i$ with $i \in \{4, 5, 6, \cdots\}$ cannot detect entanglement.
	
	(2) When $ 0.57 < p \le 0.80 $, each of the two pairs Alice$^1$-Bob$^1$ and Alice$^2$-Bob$^2$ can detect the entanglement.  Other pairs Alice$^i$-Bob$^i$ with $i \in \{3, 4, 5, \cdots\}$ cannot detect entanglement.
	
	(3) When $ 0.33< p \le 0.57$, only the pair Alice$^1$-Bob$^1$ can detect the entanglement.  Other pairs Alice$^i$-Bob$^i$ with $i \in \{2, 3, 4, \cdots\}$ cannot detect entanglement.
	
	\subsection{Initially shared non-maximally entangled two-qubit pure state }
	Suppose that Alice$^1$ and Bob$^1$ initially share a non-maximally entangled two-qubit pure state given by,
	\begin{equation}
	| \Psi \rangle = \cos \theta |01\rangle + \sin \theta |10 \rangle, 
	\end{equation}
	with $0 < \theta < \frac{\pi}{4}$. For this state also, the optimal entanglement witness operator is given by Eq.(\ref{eq1})  \cite{ewo}.
	
	In this case too, at most three pairs of Alice and Bob (e.g., Alice$^1$-Bob$^1$, Alice$^2$-Bob$^2$, Alice$^3$-Bob$^3$) can witness entanglement sequentially through the entanglement witness operator (\ref{unwo}). Further, the following results are obtained.
	
	(1) When $ \frac{\pi}{8} \leq \theta < \frac{\pi}{4}$,  each of the pairs Alice$^1$-Bob$^1$, Alice$^2$-Bob$^2$ and Alice$^3$-Bob$^3$ can detect the entanglement.  Other pairs Alice$^i$-Bob$^i$ with $i \in \{4, 5, 6, \cdots\}$ cannot detect entanglement.
	
	(2) When $ \frac{\pi}{17} \le \theta < \frac{\pi}{8} $, each of the two pairs Alice$^1$-Bob$^1$ and Alice$^2$-Bob$^2$ can detect the entanglement.  Other pairs Alice$^i$-Bob$^i$ with $i \in \{3, 4, 5, \cdots\}$ cannot detect entanglement.
	
	(3) When $ 0 < \theta < \frac{\pi}{17} $, only the pair Alice$^1$-Bob$^1$ can detect the entanglement.  Other pairs Alice$^i$-Bob$^i$ with $i \in \{2, 3, 4, \cdots\}$ cannot detect entanglement.

\subsection{Advantage of the sequential measurement  scenario} For sequential detection of entanglement by multiple pairs of observers, it is important to ensure that the expectation values of the witness operator for different pairs become as much negative as possible, for feasibility of  practical detection of entanglement. In our case, as we discussed earlier, maximum three pairs of Alice and Bob can detect entanglement. Hence, for our purpose, we define   `Detectability' ($D$) as the minus one times the sum of the expectation values of the entanglement witness operators for all the three pairs.

			Mathematically, detectability  is defined as
	\begin{align}
D= (-1) \sum_{i=1}^{3} D_{ii} &=(-1)\sum_{i}  \text{Tr}\Big[W^{(\xi_{i},\lambda_{i})} \, \rho_{A_{i}B_{i}} \Big] \nonumber \\
&\hspace{0.3cm} \text{with} \hspace{0.3cm} \text{Tr}\Big[W^{(\xi_{i},\lambda_{i})} \, \rho_{A_{i}B_{i}} \Big] < 0 \, \, \forall i,
\end{align}
where $\rho_{A_i B_i}$ is the state shared by the pair Alice$^i$-Bob$^i$.  Since, negative expectation value of the witness operator implies detection of entanglement, we have taken minus sign in the above definition to make $D$ positive when each pair detects entanglement.

Now it is of practical demand to look for a measurement strategy that would yield optimum witness of entanglement in the sequential measurement scenario. For that we need to define the maximum detectability,   $D_{\text{max}}$ which is obtained by maximizing $D$ over all possible sharpness parameters ($\xi_{i},\lambda_{i} $) of all the three pairs of observers under the constraint that each of the three pairs can detect entanglement, i.e., $ \text{Tr}\Big[W^{(\xi_{i},\lambda_{i})} \, \rho_{A_{i}B_{i}} \Big] < 0 \, \, \forall i \in \{1, 2, 3\}. $

Mathematically, maximum detectability $D_{\text{max}}$ is defined as,
\begin{align}
&D_{\text{max}} = \max_{\xi_{1}, \lambda_{1}, \xi_{2}, \lambda_{2}, \xi_{3}, \lambda_{3}} D \nonumber \\	
& \text{such that}  \nonumber \\
& D_{ii} = \text{Tr}\Big[W^{(\xi_{i},\lambda_{i})} \, \rho_{A_{i}B_{i}} \Big] < 0 \, \, \forall i \in \{1, 2, 3\}.
\end{align}	  $D_{\text{max}}$ serves as a tool to quantify the  overall ability of the three pairs of observers with their best measurement strategy to detect entanglement in an experiment in a sequential measurement scenario.   Here, the best measurement strategy implies the one that makes the expectation values of the witness operators as much negative as possible for all the  three pairs simultaneously. Note that when $D$ in the above definition is maximized, individual $\text{Tr}\Big[W^{(\xi_{i},\lambda_{i})} \, \rho_{A_{i}B_{i}} \Big]$ may not be optimized. This is because the expectation values of the witness operators of all the three pairs are not optimized simultaneously. Also, for example, when the expectation value of the witness operator for the first pair of Alice and Bob is optimized, other subsequent pairs may not detect any entanglement. Since our objective in the present paper is to optimize the expectation values of the witness operators for all the three pairs simultaneously, we have taken the above definition of maximum detectability. 

Also note here that although the above definition is expressed for three pairs of Alice and Bob (relevant for the present study), it can be generalized to any number of pairs of observers depending on the specific context under consideration.

We now demonstrate the advantage of the sequential scenario when Alice$^1$ and Bob$^1$ initially share the Bell state $\rho_{A_1 B_1} = |\psi^{+}\rangle \langle \psi^{+}|$ with  $|\psi^{+}\rangle = (|01\rangle + |10\rangle)/\sqrt{2}$. In this case, at most three pairs can detect entanglement. Hence, $D$ in this case is given by,
\begin{eqnarray}
D= &&  -D_{11}-D_{22}-D_{33} \nonumber \\
= && (-1)\sum_{i=1}^{3}  \text{Tr}\Big[W^{(\xi_{i},\lambda_{i})} \, \rho_{A_{i}B_{i}} \Big],
\end{eqnarray}
where $\rho_{A_{i}B_{i}} $ is the state shared, on average, by the pair Alice$^i$-Bob$^i$.
Based on the analysis described in Sec. \ref{sec4a}, it can be shown that
	\begin{eqnarray}
	D = && -\frac{1}{4}(1-3\xi_{1}\lambda_{1}) -\frac{1}{12} \Big[3-\Big(1+2\sqrt{1-\xi_{1}^{2}} \Big) \Big(1+2\sqrt{1-\lambda_{1}^{2}} \Big)\xi_{2}\lambda_{2} \Big] \nonumber \\
	&& - \frac{1}{108} \Big[27- \Big(1+2\sqrt{1-\xi_{1}^{2}} \Big) \Big( 1+2\sqrt{1-\lambda_{1}^{2}} \Big) \nonumber \\
	&& \hspace{1.9cm} \Big( 1+2\sqrt{1-\xi_{2}^{2}} \Big) \Big( 1+2\sqrt{1-\lambda_{2}^{2}} \Big) \xi_{3}\lambda_{3} \Big]. 
	\end{eqnarray}
Now, we get   $D_{\text{max}}=0.20$, which
is obtained for $ \xi_{1}=\lambda_{1}=0.73$, $\xi_{2}=\lambda_{2}=0.80$, and $\xi_{3}=\lambda_{3}=1$.  

	Next, let us evaluate the total RoM that is needed to achieve the above-mentioned $D_{\text{max}}$. As mentioned earlier, for the unsharp measurement $E^{\lambda}_{\hat{n}} \equiv \{E^{\lambda}_{+|\hat{n}}, E^{\lambda}_{-|\hat{n}}\}$ defined by Eq.(\ref{unsharpm}), we have  $R(M)=\lambda$.  Hence, the total RoM, denoted by $R_{\text{total}}(M)$, needed to achieve the above-mentioned $D_{\text{max}}$ is given by,
\begin{align}
&R_{\text{total}}(M) = \sum_{i=1}^{3} (\xi_i + \lambda_i) \nonumber \\ 
& \text{such that} \nonumber \\
&D=  D_{\text{max}} = 0.20.
\end{align}
Consequently, we have that
\begin{align}
R_{\text{total}}(M)&=2(0.73+0.80+1) \nonumber \\
& = 5.06
\end{align} 
$R_{\text{total}}(M)$ quantifies the total amount of resource consumed while performing the unsharp measurements necessary for achieving the maximum detectability.

	Next, we will compare the resource requirement in terms of the entanglement consumed  and total RoM between the sequential  measurement  scenario (\ref{sce1}) and the corresponding non-sequential  measurement scenario. The non-sequential measurement  scenario involves three pairs of observers- Alice$^1$-Bob$^1$, Alice$^2$-Bob$^2$ and Alice$^3$-Bob$^3$, where each of these three pairs share one copy of a two-qubit entangled state. Hence, this non-sequential  measurement scenario involves total three  pairs of entangled qubits. Each pair detects entanglement of their shared state using the witness operator (\ref{unwo}). 
For the purpose of meaningful comparison of the sequential and the non-sequential scenarios, we need to ensure that the detectability $D$ remains
the same for both the scenarios.  This will confirm that the overall ability of all the three pairs of observers to detect entanglement is the same in both the scenarios. Next, we will perform the aforementioned comparison when different
	types of entangled states are shared between the three pairs of Alice and Bob in the non-sequential scenario.

	\subsubsection{Comparison with non-sequential scenario involving three pairs of entangled qubits in  Werner states}

 Let us first consider that the  following  Werner state,

\begin{equation}
\rho_{\text{W}}(p_{i})=p_{i} |\psi^{+}\rangle \langle \psi^{+}| + (1-p_{i}) \frac{\mathbb{I}}{2} \otimes \frac{\mathbb{I}}{2}
\label{werner}
\end{equation} 
with $0 < p _{i}\leq 1$ is shared between the pair Alice$^i$-Bob$^i$ in the aforementioned non-sequential scenario with $ i\in \{1,2,3\}$. Here, $p_1,$ $p_2,$ and $p_3$ are different in general. Let $\tilde{\xi_{i}}$ and $\tilde{\lambda_{i}}$ denote the sharpness parameters for the measurements by Alice$^{i}$ and Bob$^{i}$ respectively in the non-sequential scenario. On the other hand, as mentioned earlier, $\xi_{i}$ and $\lambda_{i}$ are the sharpness parameters for the measurements by Alice$^{i}$ and Bob$^{i}$ respectively in the  sequential measurement case.  Here $\tilde{\xi_{i}}$ may or may not be equal to $\xi_{i}$ and $\tilde{\lambda_{i}}$ may or may not be equal to  $\lambda_{i}$ for all $ i\in \{1,2,3\}$ in general.

Now, we would like to evaluate the minimum amount of the total entanglement (denoted by $\eta$) that is necessary in the non-sequential measurement scenario for the detectability in the non-sequential  measurement scenario given by, $D^{\text{NS}} = (-1)\sum_{i=1}^{3} \text{Tr}\Big[W^{(\tilde{\xi_{i}},\tilde{\lambda_{i}})} \, \rho_{\text{W}}(p_{i}) \Big]=(-1)\sum_{i=1}^{3}\frac{1}{4}(1-3p_{i}\tilde{\xi_{i}}\tilde{\lambda_{i}})$ being equal to the maximum detectability in the sequential  measurement scenario denoted by $ D^{\text{S}}_{\text{max}}=0.2$ and the total RoM in the non-sequential  measurement case  given by $\sum_{i=1}^{3} (\tilde{\xi_i} +\tilde{\lambda_i})$ being equal to that in the sequential  measurement case, i.e., equal to 5.06.  Also, we must ensure that each of the pairs of Alice and Bob can detect entanglement in the non-sequential scenario while performing the above minimization problem. This last constraint is a natural demand for a meaningful comparison.

The total concurrence of the three copies of the Werner states (\ref{werner}) is, $\eta=\sum_{i=1}^{3} C(\rho_{\text{W}}(p_{i}))=\sum_{i=1}^{3} \frac{1}{2}(3p_{i}-1).$ Thus, now the task is to minimize $\eta$ with the following constraints: $ \sum_{i=1}^{3} (\tilde{\xi_i} +\tilde{\lambda_i})=5.06,$ $\sum_{i=1}^{3}\frac{1}{4}(1-3p_{i}\tilde{\xi_{i}}\tilde{\lambda_{i}})=-0.2$ and $ \frac{1}{4}(1-3p_{i}\tilde{\xi_{i}}\tilde{\lambda_{i}}) < 0 \hskip .2cm \forall \hskip .2cm i \in \{1,2,3\}$. Now it turns out that the minimum of $\eta$ under the above constraints is given by,  $\eta_{\text{min}}=1.11$.  This is achieved when $p_1 = 0.54$, $p_2 = 0.54$, $p_3 = 0.65$, $\tilde{\xi_1} =\tilde{\lambda_1}= 0.79$, $\tilde{\xi_2} =\tilde{\lambda_2} =0.79$, $\tilde{\xi_3} =\tilde{\lambda_3} =0.95$.  Now, as mentioned earlier, $D^{\text{S}}_{\text{max}}$ in the sequential scenario is achieved for
$ \xi_{1}=\lambda_{1}=0.73 $, $\xi_{2}=\lambda_{2}=0.80$,  $\xi_{3}=\lambda_{3}=1$. Hence, we don't need to take equal sharpness parameter in both the scenarios for each of the observers in order to satisfy equal amount of the total RoM in both scenarios and $D^{\text{NS}} = D^{\text{S}}_{\text{max}}$.

On the other hand, only $1$ ebit is sufficient in the sequential  measurement scenario for achieving $ D^{\text{S}}_{\text{max}}$ as only one copy of the maximally entangled state is involved. Hence, for achieving the same amount of detectability in the sequential and the non-sequential  measurement scenarios with the total resourcefulness of the measurements being equal in both  the scenarios, the non-sequential  measurement scenario needs greater amount of entanglement compared to the sequential  measurement scenario. This result is summarized in Table \ref{tab1}. This demonstrates a resource theoretic advantage of the sequential  measurement scenario over the non-sequential  measurement scenario. 

\begin{table}
	\centering
	\begin{tabular}{|*{4}{c|}}
		\hline
		\multicolumn{2}{|c|}{}	& Sequential
		& Non-sequential  \\
		\multicolumn{2}{|c|}{}	& measurement & measurement \\
		\multicolumn{2}{|c|}{}	& 
		scenario & scenario  \\
		\hline 
		\multicolumn{2}{|c|}{Detectability $D$} &  $ 0.20$ & $ 0.20$ \\ 
		\hline 
		\multicolumn{2}{|c|}{Total robustness  of} & &  \\ 
		\multicolumn{2}{|c|}{measurement $R_{\text{total}}(M)$} & $5.06$ & $5.06$ \\ 
		\hline 
		Total & with $\rho_{W}(p_{i})$ in non-sequential case	& $1$ ebit & $\ge 1.11$ ebits \\ 
		\cline{2-4} 
		entanglement & with $\rho_{p_{i}}$ in non-sequential case	& $1$ ebit & $\ge 1.11$ ebits \\ 
		\cline{2-4} 
		consumed $ \eta $	& with $|\Psi(\theta_{i})\rangle$ in non-sequential case & $1$ ebit & $\ge 1.11$ ebits\\ 
		\hline 
	\end{tabular}
	
	\caption{The total amount of required entanglement $\eta$ when each party performs equally resourceful measurements in both scenarios. Here, the non-sequential  measurement scenario involves either  three pairs of entangled qubits in the Werner states $\rho_W(p_{i})$, or three pairs of entangled qubits in the mixed states with colored noise $\rho_{p_{i}}$, or three pairs of qubits in non-maximally entangled pure states $|\Psi(\theta_{i})\rangle$. All numerical values presented in this table are rounded to two decimal places.}
	\label{tab1}
\end{table}

It is  possible to demonstrate further the advantage of the sequential  measurement scenario from another perspective as well. Let us now take the total entanglement consumed $ \eta $ in the two scenarios to be equal. The detectability $D^{\text{NS}}$  in the non-sequential  measurement scenario  is also
taken  to be equal to the maximum detectability $D^{\text{S}}_{\text{max}}$ in the sequential  measurement scenario.  We also ensure that each of the pairs of Alice and Bob can detect entanglement in the non-sequential scenario.  Under these conditions, we  compare the total RoM, i.e., $R_{\text{total}}(M)$ in the two scenarios. 

Total entanglement consumption in the sequential  measurement scenario starting with a maximally entangled initial state is $\eta$ = $1$ ebit. Hence, for the non-sequential  measurement scenario we use three different Werner states, $ \rho_{\text{W}}(p_{i}), i=1,2,3 $  with the total amount of entanglement being equal to that in the sequential  measurement scenario, i.e.,  $\sum_{i=1}^{3} C(\rho_{\text{W}}(p_{i}))=1, \hskip .2cm \text{i.e.,} \hskip .2cm p_{1}+p_{2}+p_{3}=1.67$. The other constraints are $D^{\text{NS}} =(-1) \sum_{i=1}^{3}  \text{Tr}\Big[W^{(\tilde{\xi_{i}},\tilde{\lambda_{i})}} \, \rho_{\text{W}}(p_{i}) \Big] = D^{\text{S}}_{\text{max}} = 0.20$ and $ \text{Tr}\Big[W^{(\tilde{\xi_{i}},\tilde{\lambda_{i}})} \, \rho_{\text{W}}(p_{i}) \Big]<0 \hskip .2cm \forall \hskip .2cm i \in \{1,2,3\} $. Now, the task is to minimize the total RoM, $R_{\text{total}}(M)= \sum_{i=1}^{3} (\tilde{\xi_i} +\tilde{\lambda_i})$ in the non-sequential scenario with the aforementioned constraints. It turns out that the minimum total RoM in the non-sequential scenario is 5.16.   

On the other hand, the total RoM in the sequential  measurement scenario is given by $5.06$, as mentioned earlier. Hence, for achieving the same amount of detectability using the same amount of total initial entanglement in the sequential and the non-sequential  measurement scenarios, the total resourcefulness of the measurements (in terms of total RoM) necessary in the non-sequential  measurement scenario 
turns out to be greater than that in the sequential  measurement scenario.
Thus a resource theoretic  advantage of the sequential  measurement scenario over the non-sequential one is clearly implied. These results are depicted in Table \ref{tab2}. 

\begin{table}
	\centering
	\begin{tabular}{|*{4}{c|}}
		\hline
		\multicolumn{2}{|c|}{}	& Sequential
		& Non-sequential  \\
		\multicolumn{2}{|c|}{}	& measurement & measurement \\
		\multicolumn{2}{|c|}{}	& 
		scenario & scenario  \\
		\hline 
		\multicolumn{2}{|c|}{Detectability $D$} &  $ 0.20 $ & $ 0.20 $ \\ 
		\hline 
		\multicolumn{2}{|c|}{Total entaglement} & &  \\ 
		\multicolumn{2}{|c|}{consumed $ \eta $} & $1$ ebit & $1$ ebit \\ 
		\hline 
		Total  & with $ \rho_{\text{W}} (p_{i}) $ in non-sequential case	& $5.06$  & $\geq 5.16$ \\ 
		\cline{2-4} 
		robustness of & with $\rho_{p_{i}}$	in non-sequential case & $5.06$ & $ \geq 5.17 $ \\ 
		\cline{2-4} 
		measurement 	& with $|\Psi(\theta_{i})\rangle$ in non-sequential case & $5.06$ & $ \geq 5.17 $\\ 
		\hline 
	\end{tabular}
	
	
	\caption{The required total robustness  of measurement $R_{\text{total}}(M)$  when the same amount of total entanglement is consumed in the both scenarios. The non-sequential  measurement scenario involves either  three pairs of entangled qubits in the Werner states $\rho_W(p_{i})$, or three pairs of entangled qubits in the mixed states with colored noise $\rho_{p_{i}}$, or three pairs of qubits in non-maximally entangled pure states $|\Psi(\theta_{i})\rangle$. All numerical values presented in the table are rounded to two decimal places.}
	\label{tab2}
\end{table}

 \subsubsection{Comparison with non-sequential scenario involving three pairs of entangled qubits in mixed states with colored noise}

Now, we consider that the pair Alice$^i$-Bob$^i$ (with $i \in \{1,2,3\}$) in the non-sequential measurement  scenario shares the following mixed state with colored noise, 
\begin{equation}
\rho_{p_{i}}=p_{i} |\phi^{+}\rangle \langle\phi^{+}|+\frac{1-p_{i}}{2}(|01\rangle \langle01|+|10\rangle \langle10|),
\label{statenew2}
\end{equation}
with $0 < p_{i} \leq 1$. 

Before proceeding, we will construct the optimal entanglement witness operator of the above state using the method described in \cite{ewo,mlew1,mlew2}. First, we  compute the eigenvector corresponding to the negative eigenvalue of $\rho_{p_{i}}^{T_B}$, where $\rho_{p_{i}}^{T_B}$ denotes the partial transposition of $\rho_{p_{i}}$. Then the entanglement witness operator is given by the partially transposed projector onto that eigenvector.  Following this approach, we obtain  the following optimal entanglement witness operator for the  state (\ref{statenew2}),
\begin{equation}\label{witness2}
\widetilde{W}=\frac{1}{4} \Big( \mathbb{I} \otimes \mathbb{I} - \sigma_{x} \otimes \sigma_{x} + \sigma_{y} \otimes \sigma_{y} - \sigma_{z} \otimes \sigma_{z} \Big).
\end{equation}	
 Now, suppose in the non-sequential measurement scenario Alice$^i$ performs unsharp measurements with sharpness parameter $\tilde{\xi_{i}}$ $\in$ $(0,1]$ and Bob$^i$ performs unsharp measurements with sharpness parameter $\tilde{\lambda_{i}}$ $\in$ $(0,1]$. Following the calculations mentioned in Sec. \ref{sec3a}, it can be shown that the modified entanglement witness operator of the state (\ref{statenew2})  for Alice$^i$ and Bob$^i$ is given by,
\begin{equation}
\widetilde{W}^{(\tilde{\xi_i},\tilde{\lambda_i})}=\frac{1}{4} \Big[\mathbb{I} \otimes \mathbb{I} - \tilde{\xi_i} \, \tilde{\lambda_i} \Big( \sigma_{x} \otimes \sigma_{x}-\sigma_{y}\otimes\sigma_{y}+\sigma_{z}\otimes \sigma_{z} \Big) \Big].
\end{equation}
Following the method mentioned in Sec. \ref{sec3a}, it can be easily shown that for any separable state $\rho_s \in \mathcal{S}$, $\text{Tr} \Big(\widetilde{W}^{(\tilde{\xi_i},\tilde{\lambda_i})} \rho_s \Big) \geq 0$ for all $\tilde{\xi_i}. \tilde{\lambda_i}$ $\in$ $(0,1]$.

 As mentioned earlier, in the sequential  measurement scenario, $D^{\text{S}}_{\text{max}}$  is achieved for $\xi_{1}=\lambda_{1}=0.73$, $\xi_{2}=\lambda_{2}=0.80$, and $\xi_{3}=\lambda_{3}=1$, i.e., total RoM=5.06. Now, we would like to evaluate what is the minimum amount of the total entanglement  $\eta$ that is necessary in the non-sequential measurement scenario for the the total RoM in the non-sequential  measurement case being equal to that in the sequential  measurement case and for the detectability in the non-sequential  measurement scenario given by, $D^{\text{NS}} = (-1)\sum_{i=1}^{3} \text{Tr}\Big[\widetilde{W}^{(\tilde{\xi}_{i},\tilde{\lambda}_{i})} \, \rho_{p_{i}} \Big]=(-1)\sum_{i=1}^{3}\frac{1}{4}\Big[1+\tilde{\xi}_{i}\tilde{\lambda}_{i}(1-4p_{i})\Big]$ being equal to  $D^{\text{S}}_{\text{max}}=0.20$.   Here also we must ensure that each pair can detect entanglement in the non-sequential scenario. The total concurrence of the three states is, $\eta=\sum_{i=1}^{3} C(\rho_{p_{i}})=\sum_{i=1}^{3} (2p_{i}-1).$ Thus, now the task is to minimize $\eta$ with the following constraints: $ \sum_{i=1}^{3} (\tilde{\xi}_i +\tilde{\lambda}_i)=5.06, \,$  $\sum_{i=1}^{3}\frac{1}{4}\Big[1+\tilde{\xi}_{i}\tilde{\lambda}_{i}(1-4p_{i})\Big]=-0.2,$ and $\frac{1}{4}\Big[1+\tilde{\xi}_{i}\tilde{\lambda}_{i}(1-4p_{i})\Big] < 0$  for all $i \in \{1,2,3\}$. Now it turns out that  $\eta_{\text{min}}=1.11.$  On the other hand, only $1$ ebit is sufficient in the sequential  measurement scenario for achieving $D^{\text{S}}_{\text{max}}$ as only one copy of the maximally entangled state is involved. So here also, the total entanglement consumption in the non-sequential scenario is greater than the sequential scenario for achieving the same detectability using equally resourceful measurements in the two scenarios. This result is summarized in Table \ref{tab1}. This again demonstrates a resource theoretic advantage of the sequential  measurement scenario over the non-sequential  measurement scenario.

Next, we take the total entanglement consumed $\eta$ in the non-sequential scenario to be equal to that in the sequential scenario (= 1 ebit) which gives rise the following constraint, $p_{1}+p_{2}+p_{3}=2$.  Also, the detectability $D^{\text{NS}}$  in the non-sequential  measurement scenario has  to be equal to the maximum detectability $D^{\text{S}}_{\text{max}}$ in the sequential measurement  scenario, i.e.,  $\sum_{i=1}^{3}\frac{1}{4}\Big[1+\tilde{\xi}_{i}\tilde{\lambda}_{i}(1-4p_{i})\Big]=-0.2$. Finally, we must ensure that each pair in the non-sequential scenario detects entanglement- $ \frac{1}{4}\Big[1+\tilde{\xi}_{i}\tilde{\lambda}_{i}(1-4p_{i})\Big] <0, \, \, \forall \, i \in \{1,2,3\}$. Under these conditions, we  compare the total RoM, i.e., $R_{\text{total}}(M)$ in the two scenarios. 
	Here too, we observe that the minimum amount of total RoM required in the non-sequential  measurement scenario is given by, $R_{\text{total}}^{\text{min}}(M) = 5.17$.   These results are depicted in Table \ref{tab2}. 	This again signifies a resource theoretic  advantage of the sequential  measurement scenario over the non-sequential one.

\subsubsection{Comparison with non-sequential scenario involving three pairs of qubits in non-maximally entangled pure states}

Next, we perform a  similar comparison between the sequential  and the non-sequential  measurement scenario,  where the pair Alice$^i$-Bob$^i$ (with $i \in \{1,2,3\}$) in the non-sequential measurement  scenario shares the following non-maximally  entangled pure state, 
\begin{equation}
| \Psi(\theta_{i}) \rangle = \cos \theta_{i} |01\rangle + \sin \theta_{i} |10 \rangle, 
\end{equation}
with $0 < \theta_{i} < \frac{\pi}{4}$. In the non-sequential  measurement scenario, each pair of Alice and Bob detects entanglement of the above pure state  through the entanglement witness operator given by Eq.(\ref{unwo}).

Similar to the earlier cases, at first, we calculate the total amount of entanglement $\eta$ that is necessary in the non-sequential  measurement scenario involving three different non-maximally entangled pure states $| \Psi(\theta_{i}) \rangle$ for the detectability in the non-sequential  measurement scenario: $D^{\text{NS}}$ $=$ $\sum_{i=1}^{3}$   $\text{Tr}$ $\Big[W^{(\tilde{\xi}_{i},\tilde{\lambda}_{i})}$  $| \Psi(\theta_{i}) \rangle \langle \Psi(\theta_{i}) | \Big]$ to be equal to the maximum detectability in the sequential  measurement scenario  and for the total RoM in the non-sequential  measurement scenario to be equal to that of the sequential scenario.  Also, we ensure that each pair in the non-sequential case detects entanglement. The results in this case can be obtained following the similar steps as adopted in the previous two cases. These results are summarized in Table \ref{tab1}. Here too, it is evident that the sequential  measurement scenario is advantageous over the non sequential scenario.

Now, let us consider that the total entanglement consumed $ \eta $ in the non-sequential  measurement scenario is equal to that in the sequential  measurement scenario (= 1 ebit). Also, consider that the detectability  in the non-sequential  measurement scenario  is equal to the maximum detectability  in the sequential  measurement scenario  and  each of the three pairs in the non-sequential case detects entanglement. Under these conditions we have again found that the minimum amount of total measurement resource  required in the non-sequential scenario is greater than that  in the sequential one  (see Table \ref{tab2}). Thus, the advantage of the sequential measurement scenario over the non-sequential one is reinforced.

 \section{Quantum teleportation}\label{teleportation} 

In this section, we illustrate the resource theoretic efficacy of the
sequential measurement scheme by presenting an example of witnessing entangled states useful for quantum teleportation.
There exist a class of Hermitian witness operators \cite{prl_nirman} that can detect entangled states useful for performing quantum teleportation \cite{teleport}.  We will show below that  a single copy of an entangled state can be recycled such that multiple pairs of observers can detect entangled states suitable for quantum teleportation. This thus demonstrates that if a pair detects entanglement using a quantum state, then the residual entanglement after this entanglement detection can again be used to perform quantum teleportation by another pair of observers. However, in the conventional approach employing non-sequential measurement strategies, more than one copies of entangled states are required for sequentially performing the two tasks- entanglement detection and quantum teleportation.

Here, we consider that the following maximally entangled mixed state \cite{mems} is initially shared in the sequential measurement scenario,
\begin{align}
\rho_{\text{MEMS}} &=\begin{pmatrix}
h(c) & 0 & 0 & \frac{c}{2} \\
0    & 1-2h(c) & 0 & 0 \\
0 & 0 & 0 & 0 \\
\frac{c}{2} & 0 & 0 & h(c)
\end{pmatrix}, \nonumber \\
&\text{with} \, \, \, h(c) = \begin{dcases}
    \frac{c}{2},& \text{if } c\geq \frac{2}{3}, \\
    \frac{1}{3},& \text{if } c < \frac{2}{3}.
\end{dcases}
\label{mems}
\end{align} 	 
 with $ c $ being the concurrence of $\rho_{\text{MEMS}}$.   The teleportation witness operator for this state is given by \cite{prl_nirman},
\begin{equation}
W_{\text{tel}}= \frac{1}{4} \Big( \mathbb{I} \otimes \mathbb{I} - \sigma_{x} \otimes \sigma_{x} + \sigma_{y} \otimes \sigma_{y} - \sigma_{z} \otimes \sigma_{z} \Big),
\label{teleportationwitness}
\end{equation} 
which satisfies the following properties: $\text{Tr}(W_{\text{tel}}\,\sigma) \ge 0$ for all states $\sigma$ that are not useful for teleportation and $\text{Tr}(W_{\text{tel}}\,\rho) < 0$ for at least one state $ \rho $ that is useful for teleportation.

Now using the same prescription described earlier  it can be shown that at most three pairs of Alice and Bob in the aforementioned sequential measurement scenario can get negative expectation values of the witness operator (\ref{teleportationwitness}) when a single copy of the state (\ref{mems}) is initially shared. Further, three pairs can have negative expectation values of the witness operator (\ref{teleportationwitness}) when $ \rho_{\text{MEMS}} $ with concurrence $ c\ge 0.93 $ is recycled.

Next, in the present context, let us compare the sequential scenario and the corresponding non-sequential scenario. Let in the sequential scenario, Alice$^1$ and Bob$^1$ initially share the state $\rho_{\text{MEMS}}$ with the minimum amount of necessary concurrence such that three pairs can get negative expectation values of the witness operator (\ref{teleportationwitness}), i.e., with concurrence $c = 0.93$.  Following the similar calculations as described earlier it can be shown that the maximum detectability in the sequential case is given by, $D^{\text{S}}_{\text{max}}=0.11$. This occurs for total RoM being equal to $4.96$.  

Next, in the  non-sequential scenario, we consider that each pair Alice$^i$-Bob$^i$ with $i \in \{1,2,3\}$ shares the state $\rho_{\text{MEMS}}$ given by (\ref{mems}) with concurrence $c_i$. Now, we compute the total amount of entanglement $\eta = c_1+c_2+c_3$ that is necessary in the non-sequential  measurement scenario for the detectability in the non-sequential  measurement scenario being equal to the maximum detectability in the sequential  measurement scenario  and for the total RoM in the non-sequential  measurement scenario to be equal to that in the sequential scenario. Also, we ensure that each pair in the non-sequential case detects useful state for teleportation. It turns out that minimum total entanglement consumed in the non-sequential scenario is 1.98 ebits, thus clearly implying an advantage of the sequential scenario.

Therefore, this analysis implies that the sequential scenario requires less resource compared to the corresponding non-sequential scenario for executing sequential detection of entangled states or useful states for quantum teleportation.

	\section{More number of observers in an asymmetric scenario}\label{sec5}
	
	In the scenario considered so far in this work, we assume that the number of sequential Alices on one side is equal to the number of sequential Bobs on the other.  We  call this scenario as a ``Symmetric Scenario''. At this stage
	it may be pertinent to ask the question as to
	what happens if we relax this condition of symmetry. To this end, let us now
	consider  sequential detection of entanglement in the case when the number of Alices is not equal in general to the number of Bobs. We consider that the Bell state given by, $|\psi^{+}\rangle = \frac{1}{\sqrt{2}} (|01\rangle + |10\rangle)$ is initially shared.
	
	It may be noted here that an asymmetric scenario involving a single Alice and multiple Bobs (Bob$^1$, Bob$^2$, Bob$^3$, $\cdots$, Bob$^m$) was discussed in \cite{bera_entanglement}. In this case, it was shown that at most twelve Bobs can detect entanglement with the single Alice \cite{bera_entanglement}.   
	
	Next, consider the scenario involving two Alices (Alice$^1$ and Alice$^2$) and multiple Bobs (Bob$^1$, Bob$^2$, Bob$^3$, $\cdots$, Bob$^m$). At first, the pair Alice$^1$-Bob$^1$ detects entanglement using the entanglement witness operator (\ref{unwo}). As mentioned in Sec. \ref{sec4a}, this pair can detect entanglement if the condition (\ref{con1}) is satisfied. The state $\rho_{A_{2}B_{2}}$ received, on average, by  Alice$^2$-Bob$^2$ from Alice$^1$-Bob$^1$ is given by Eq.(\ref{statea2b2}). 
	Since, there are only two Alices in the sequence, Alice$^2$  performs projective measurements. On the other hand, all subsequent Bobs (i.e., Bob$^2$, Bob$^3$, $\cdots$, Bob$^m$)  perform unsharp measurements. The sharpness parameter associated with the  measurement by Bob$^i$ (with $i \in \{2, 3, \cdots, m\}$) is denoted by $\lambda_i$. Hence, the modified entanglement witness operator used by the pair Alice$^2$-Bob$^i$ is given by, 
	\begin{align}
	W^{(\lambda_i)} =  \frac{1}{4} \Big[ \mathbb{I} \otimes \mathbb{I} +  \lambda_i \Big( \sigma_{z} \otimes  \sigma_{z} -  \sigma_{x} \otimes  \sigma_{x}  -  \sigma_{y} \otimes   \sigma_{y} \Big) \Big]. 
	\label{unwoasy}
	\end{align}	
	This can be obtained from the entanglement witness operator given by Eq.(\ref{eq1}) and following the analysis mentioned in Sec. \ref{sec3a}.  It can be shown that for any separable state $\rho_s \in \mathcal{S}$, we have $\text{Tr} \Big(W^{( \lambda_j)} \rho_s \Big) \geq 0$.	
	
	Alice$^2$ and Bob$^2$ can witness entanglement if
	\begin{equation}
	\text{Tr} \Big[W^{(\lambda_{2})} \, \rho_{A_{2}B_{2}} \Big]<0, 
	\end{equation}
	which implies the following condition,	
	\begin{equation}
	\lambda_{2}>\frac{3}{ \Big(1+2\sqrt{1-\xi_{1}^{2}} \Big) \Big(1+2\sqrt{1-\lambda_{1}^{2}} \Big)}.
	\label{cond1asy}
	\end{equation}
	The average state shared between Alice$^2$ and Bob$^3$ is given by,
	\begin{align}
	\rho_{A_{2}B_{3}} =\frac{1}{3}\sum_{m_2,b_2} \Big( \mathbb{I} \otimes \sqrt{E^{\lambda_{2}}_{b_2|\hat{m}_2}} \Big) \, \rho_{A_{2}B_{2}} \,  \Big( \mathbb{I} \otimes \sqrt{E^{\lambda_{2}}_{b_2|\hat{m}_2}} \Big),
	\label{poststa2b3}
	\end{align}
	with	$\hat{m}_2 \in \{ \hat{x}, \hat{y}, \hat{z}\}$ and $b_2 \in \{+1,-1\}$. Here also, we  use the assumption  that all possible measurement settings of Bob$^2$ are equally probable. 
	
	Similarly, Alice$^2$ and Bob$^3$ can witness entanglement if the following condition is satisfied,
	\begin{equation}
	\text{Tr} \Big[W^{(\lambda_{3})} \, \rho_{A_{2}B_{3}} \Big]<0.
	\end{equation}
	The above condition implies the following,
	\begin{equation}
	\lambda_{3}>\frac{9}{ \Big(1+2\sqrt{1-\xi_{1}^{2}} \Big) \Big(1+2\sqrt{1-\lambda_{1}^{2}} \Big) \Big(1+2\sqrt{1-\lambda_{2}^{2}} \Big)}.
	\label{cond2asy}
	\end{equation}
	Repeating the above calculations for other subsequent Bobs (i.e., Bob$^4$, Bob$^5$, $\cdots$, Bob$^m$), we can derive the conditions on $\lambda_4$, $\lambda_5$, $\cdots$.
	
	Next,  we determine what is the maximum number of Bobs succeed in witnessing the entanglement of the shared state with Alice$^2$. Combining the conditions (\ref{con1}), (\ref{cond1asy}), (\ref{cond2asy}), other similar conditions on $\lambda_4$, $\lambda_5$, $\cdots$, and performing  analytical calculations as described in  Appendix \ref{a1}, we get that Bob$^2$, Bob$^3$, $\cdots$, Bob$^8$ can  detect entanglement with Alice$^2$, if the following conditions are satisfied simultaneously,
	\begin{align}
	& \xi_1 \,= \lambda_1 = 0.58 + \tilde{\delta_1}  \, \, \text{with} \, \, 0 \le \tilde{\delta_1} <<1,\nonumber \\
	& \lambda_2 = 0.44 + \tilde{\delta_2} \, \, \text{with} \, \, 0 \le \tilde{\delta_2} <<1,\nonumber \\
	&  \lambda_3 = 0.47 + \tilde{\delta_3}  \, \, \text{with} \, \, 0 \le \tilde{\delta_3}<<1,\nonumber \\
	&  \lambda_4 = 0.51 + \tilde{\delta_4}  \, \, \text{with} \, \, 0 \le \tilde{\delta_4} <<1,\nonumber \\
	&  \lambda_5 = 0.56 + \tilde{\delta_5}  \, \, \text{with} \, \, 0 \le \tilde{\delta_5} <<1,\nonumber \\
	&  \lambda_6 = 0.63 + \tilde{\delta_6}  \, \, \text{with} \, \, 0 \le \tilde{\delta_6}<<1,\nonumber \\
	&  \lambda_7 = 0.74 + \tilde{\delta_7}  \, \, \text{with} \, \, 0 \le \tilde{\delta_7}<<1,\nonumber \\
	&  \lambda_8 = 0.95 + \tilde{\delta_8}  \, \, \text{with} \, \, 0 \le \tilde{\delta_8}<<1.
	\label{allcondasy}
	\end{align}
	Here the numerical values appearing  in  the above conditions are rounded to two decimal places.

	If Alice$^1$, Alice$^2$, Bob$^1$, Bob$^2$, Bob$^3$, $\cdots$, Bob$^8$ perform particular measurements with sharpness parameters satisfying the conditions mentioned in (\ref{allcondasy}), then it can be shown that Bob$^9$ cannot witness entanglement with Alice$^2$ even if Bob$^9$ performs projective measurements, i.e., with $\lambda_{9}=1$. Hence, at most eight sequential Bobs can detect entanglement in this scenario with two Alices. 
	
	Next, let us consider the scenario involving three Alices (Alice$^1$, Alice$^2$ and Alice$^3$) and multiple Bobs (Bob$^1$, Bob$^2$, Bob$^3$, $\cdots$, Bob$^m$). At first, the pair Alice$^1$-Bob$^1$ detects entanglement using the entanglement witness operator (\ref{unwo}). Then Alice$^1$ and Bob$^1$ pass their particles to Alice$^2$ and Bob$^2$, respectively. The pair Alice$^2$-Bob$^2$ detects entanglement using the same entanglement witness operator (\ref{unwo}) and passes the respective particles to the pair Alice$^3$-Bob$^3$ who performs measurements
	to detect entanglement. Now, Bob$^3$ passes his particle to Bob$^4$ so that the pair Alice$^3$-Bob$^4$ can detect entanglement. Subsequently,  Bob$^4$  passes the particle to Bob$^5$, and so on.  In this scenario, following the aforementioned calculations, it can be shown that Bob$^3$, Bob$^4$ and Bob$^5$ can detect entanglement with Alice$^3$. No additional Bob can detect entanglement with Alice$^3$. Hence, at most five Bobs can detect entanglement.

	Finally, if we consider that  there are four Alices (Alice$^1$, Alice$^2$, Alice$^3$ and Alice$^4$) and multiple Bobs (Bob$^1$, Bob$^2$, Bob$^3$, $\cdots$, Bob$^m$), then the result derived in Sec. \ref{sec4a} for the symmetric scenario tells us that it is not possible for four pairs (i.e., Alice$^1$-Bob$^1$, Alice$^2$-Bob$^2$, Alice$^3$-Bob$^3$ and Alice$^4$-Bob$^4$) to detect entanglement sequentially. Entanglement detection is possible only up to the third pair. 
	
	Before concluding, it may be noted that a resource theoretic comparison of
	the above asymmetric scenarios can be performed with the corresponding
	non-sequential scenarios involving multiple copies of mixed or non-maximally
	entangled  pure initial states. As expected, similar to the case of the symmetric scenario, here
	too it is possible to observe advantages of the sequential scenario in terms of
	the entanglement consumed and robustness of measurement.

	\section{Concluding Discussions} \label{sec6}
	
	To summarize,  our analysis presented in this paper clearly shows a
	hitherto unexplored resource theoretic advantage of recycling  a single copy of a two-qubit quantum
	state towards detecting entanglement by multiple observers. Specifically,
	we have analyzed in detail a scenario involving multiple independent observers acting 
	sequentially on each of the two spatially separated wings that
	initially share the resource of a single copy of a two-qubit entangled state. The number
	of observers on the two wings may be equal or unequal, depending upon the type
	of network considered. We have estimated the maximum number of observers that can detect entanglement sequentially using only one pair of qubits.  Our results indicate that one can reduce the number of physical qubits needed in the context of performing different entanglement-assisted quantum tasks  multiple times in network scenarios.
	
	Furthermore,  we have performed a quantitative analysis of the advantage of the sequential measurement scenario involving multiple sequential observers on both wings from the perspective of resource requirements. In particular, we have shown that the above scenario can help in reducing the necessary requirement of the total initial entanglement as well as the resource cost of the measurements 
	necessary for detecting entanglement by multiple sequential observers.  These advantages demonstrate the benefits of recycling single-shot entanglement  in various entanglement-assisted quantum information processing and communication tasks in practical contexts, in comparison with various standard schemes
	employing either multiple copies of two-qubit entangled states, or multipartite
	entangled states for performing such tasks.

	The analysis of this paper may be extended to certain  interesting
	directions. The scheme of unsharp measurement  that we have adopted in this work employs the same value of the sharpness parameter associated with  measurements in all directions by an individual
	observer.  Considering different sharpness 
	parameters for different measurement input settings by an observer can lead to a 
	significant increase in the number of allowed observers \cite{prl_brown,bntwo,cabello}. The resource theoretic efficacy of such a scheme would then be worthwhile
	to study. A fertile direction of study could also be to generalize the
	scheme formulated in the present paper towards studying resource theoretic efficacy of sequential sharing
	of single-shot entanglement in the context of multi-qubit  and two-qudit
	higher dimensional  entangled states. Finally, categorizing various information
	processing tasks involving sequential measurements \cite{rantt,cctt,appln1tt,ractt,appln4tt,appln5tt,sroytt,rsp_shounak,rac_debarshi} in terms of their resource theoretic advantages is
	another potentially attractive direction of future study.

Before concluding,  it may be noted that  several entanglement witnessing based tasks (for example, detecting eavesdropping in quantum key distribution \cite{ewapp1}, estimating localizable entanglement \cite{ewapp2})  may be required to be performed sequentially in order to execute some communication/computation  protocol in practical situations.  This can be achieved in one of the following two ways: (i) by using different copies of quantum states for different rounds of tasks, or (ii) by recycling the same copy of a quantum states. Obviously, the second approach requires less number of physical systems. However, from a resource theoretic point of view, it had been  unclear before the analysis of our present work whether the second method really consumes less resource (either in terms of total entanglement, or in terms of total resourcefulness of all the measurements performed). This is because it may so happen that the total amount of entanglement of all quantum states necessary in the first approach is smaller than the necessary entanglement of the single copy of the quantum state needed in the second approach. It is this void in the literature that the present paper seeks to fill in by establishing a resource centric advantage of the second method. However, the present study does not capture the advantage of the sequential scenario from all perspectives. For example, sharing one copy of an entangled state and sharing multiple copies of entangled states between spatially separated observers require different experimental efforts. Therefore, probing a more general resource based comparison between sequential and non-sequential scenario incorporating all such factors related to practical implementations is worth for future research and our present analysis
	indeed motivates  future studies along this direction.

	\vskip 0.5in

	\section*{Acknowledgement}
	AKD, DH, ASM acknowledge support from the project no.\\ DST/ICPS/QuEST/2018/98 of the Department of Science and Technology, Government of India. AKD acknowledges Shashank Gupta for useful discussions. DD acknowledges  Science and Engineering Research Board (SERB), Government of India for financial support through a National Post Doctoral Fellowship (File No.: PDF/2020/001358). During the later phase of this work, the research of DD has been supported by the Royal Society (United Kingdom) through the Newton International Fellowship (NIF$\backslash$R$1\backslash212007$).  SM acknowledges the Ministry of Science and Technology, Taiwan (Grant No. MOST 111-2124-M-002-013). DH acknowledges support from the   NASI Senior Scientist
	Fellowship.
	
	\section*{Data availability statement}
	
	Data sharing is  not applicable to this article as no datasets were generated or analysed during the current study.

	\appendix

		\section{Appendix} \label{a1}
		
		The condition for Alice$^1$ and Bob$^1$ to witness entanglement is given by,  	
		\begin{equation}
		\xi_{1} \, \lambda_{1} > \frac{1}{3} 
		\label{appe1}
		\end{equation}
		Similarly, the condition for Alice$^2$ and Bob$^2$ to witness  entanglement is given by,  
		\begin{equation}
		\xi_{2} \, \lambda_{2} > \frac{3}{\Big(1+2\sqrt{1-\xi_{1}^{2}} \Big) \Big(1+2\sqrt{1-\lambda_{1}^{2}} \Big)}.
		\label{appe2}
		\end{equation}
		Now, in order to ensure maximum number of sequential observers witnessing entanglement, Alice$^1$ and Bob$^1$ should choose the sharpness parameters of their measurements in such a way that these can detect entanglement causing minimal disturbance to the state. In any unsharp measurement of the form (\ref{unsharpm}) on qubits, the disturbance can be reduced by reducing the associated sharpness parameter \cite{mathematics_mal,sasmal_steering}.   Hence, Alice$^1$ and Bob$^1$ should choose sharpness parameters satisfying the following relation,
		\begin{equation}
		\xi_{1} \, \lambda_{1}  = \frac{1}{3} + \epsilon_1 \, \, \, \text{with} \, \, \, 0 < \epsilon_1 <<1.
		\label{appe3}
		\end{equation}

		Next, 	the sharpness parameters of the measurements by Alice$^2$ and Bob$^2$ also should ensure  detection of entanglement causing minimal disturbance to the state. For this, we have to minimize the right hand side of (\ref{appe2}) under the constraint given by Eq.(\ref{appe3}).	 In other words, we have to perform the following optimization problem,
	\begin{align}
		&\max_{\xi_1,\lambda_1} f_1( \xi_1, \lambda_1)\nonumber \\	
		& \text{such that}  \nonumber \\
		&\xi_{1} \, \lambda_{1}  = \frac{1}{3} + \epsilon_1 \nonumber \\
		&0 < \epsilon_1 <<1, \nonumber \\
		& 0 < \xi_1, \lambda_1 \leq 1, 
		\end{align}
		where
		\begin{align}
		   f_1( \xi_1, \lambda_1) =\Big(1+2\sqrt{1-\xi_{1}^{2}} \Big) \Big(1+2\sqrt{1-\lambda_{1}^{2}} \Big).
		\end{align} 
		
		Now, taking  $\epsilon_1 = 10^{-2}$, we obtain $\xi_1 =\lambda_1= 0.58$. Hence, the condition (\ref{appe2}) becomes 
		\begin{equation}
		\xi_2 \, \lambda_2 > 0.43.
		\end{equation}
		(Note that all numerical values appearing in this appendix are rounded to two decimal places.)
		
		Next, we find out the conditions under which Alice$^1$-Bob$^1$, Alice$^2$-Bob$^2$ and Alice$^3$-Bob$^3$ can witness entanglement in such a way that the measurement by each observer causes minimum possible disturbance to the state.
		Alice$^3$-Bob$^3$ can detect entanglement if  
		\begin{equation}
		\xi_{3} \, \lambda_{3} >\frac{27}{\prod\limits_{i=1}^{2} \left[\Big(1+2\sqrt{1-\xi_{i}^{2}} \Big) \Big(1+2\sqrt{1-\lambda_{i}^{2}} \Big) \right]}.
		\label{con3app}
		\end{equation} 
		
		In order to ensure minimum possible disturbance by Alice$^3$-Bob$^3$ while winessing entanglement, we  minimize the right hand side of (\ref{con3app}) performing the following optimization problem,
	\begin{align}
		&\max_{\xi_2,\lambda_2} f_2( \xi_2, \lambda_2)  \nonumber \\	
		& \text{such that} \nonumber \\
		&\xi_{2} \, \lambda_{2}  =0.43 + \epsilon_2 , \nonumber \\ 
		& 0 < \epsilon_2 <<1, \nonumber \\
		& 0 < \xi_2, \lambda_2 \leq 1 ,
		\end{align}
		where
		\begin{align} 
		f_2( \xi_2, \lambda_2) =\prod\limits_{i=1}^{2} \left[\Big(1+2\sqrt{1-\xi_{i}^{2}} \Big) \Big(1+2\sqrt{1-\lambda_{i}^{2}} \Big) \right] \hspace{0.4cm} \text{with} \hspace{0.4cm} \xi_1 = \lambda_1 = 0.58.
		\end{align} 
		Taking $\epsilon_2 = 10^{-2}$, we obtain  $\xi_2  = \lambda_2= 0.66$. With these values, the condition (\ref{con3app}) becomes
		\begin{equation}
		\xi_{3} \, \lambda_{3} > 0.62.
		\end{equation}
		
		Proceeding in a similar way, we check the conditions on the parameters under which Alice$^1$-Bob$^1$, Alice$^2$-Bob$^2$, Alice$^3$-Bob$^3$ and Alice$^4$-Bob$^4$ can witness entanglement.  Alice$^4$-Bob$^4$ can detect entanglement if	\begin{equation}
		\xi_{4} \, \lambda_{4} > \frac{243}{ \prod\limits_{i=1}^{3} \left[\Big(1+2\sqrt{1-\xi_{i}^{2}} \Big) \Big(1+2\sqrt{1-\lambda_{i}^{2}} \Big) \right]}.
		\label{con4app}
		\end{equation} 
		
		Here the corresponding optimization problem is,
		\begin{align}
		&\max_{\xi_3,\lambda_3} f_3( \xi_3, \lambda_3) \nonumber \\	
		& \text{such that} \nonumber \\	
		&\xi_{3} \, \lambda_{3}  =0.62 + \epsilon_3 , \nonumber \\ 
		&0 < \epsilon_3 <<1, \nonumber \\
		&0 < \xi_3, \lambda_3 \leq 1,
		\end{align}
		where
		\begin{align} 
		& f_3( \xi_3, \lambda_3) = \prod\limits_{i=1}^{3} \left[\Big(1+2\sqrt{1-\xi_{i}^{2}} \Big) \Big(1+2\sqrt{1-\lambda_{i}^{2}} \Big) \right] \nonumber \\
		&\hspace{5.5cm} \text{with} \hspace{0.4cm} \xi_1 = \lambda_1 = 0.58, \, \, \xi_2 = \lambda_2 = 0.66.
		\end{align} 
		Taking  $\epsilon_3 = 10^{-2}$, we get $\xi_3 = \lambda_3 = 0.79$. 
		However, with these values, Eq.(\ref{con4app}) becomes 
		\begin{equation}
		\xi_{4} \, \lambda_{4} > 1.13.
		\end{equation}
		Since $\xi_{4}, \lambda_{4} \in (0,1]$, the above condition cannot be satisfied. 
		Therefore, at most three pairs (Alice$^1$-Bob$^1$, Alice$^2$-Bob$^2$, Alice$^3$-Bob$^3$) can detect entanglement.


\begin{thebibliography}{99}
		\bibitem{ent1} E. Schrödinger, \emph{Die gegenwartige Situation in der Quantenmechanik}, \href{https://doi.org/10.1007/BF01491891}{Naturwissenschaften {\bf 23}, 807 (1935).}
		
		\bibitem{ent2} A. Einstein, B. Podolsky, and N. Rosen, \emph{Can Quantum-Mechanical Description of Physical Reality Be Considered Complete?}, \href{https://journals.aps.org/pr/abstract/10.1103/PhysRev.47.777}{Phys. Rev. {\bf 47}, 777
			(1935).}
		
		\bibitem{ent3} R. Horodecki, P. Horodecki, M. Horodecki, and K. Horodecki, \emph{Quantum entanglement}, \href{https://journals.aps.org/rmp/abstract/10.1103/RevModPhys.81.865}{Rev. Mod. Phys. {\bf 81}, 865 (2009).}
		
		\bibitem{belln1} J. S. Bell, \emph{On the Einstein Podolsky Rosen paradox}, \href{https://journals.aps.org/ppf/abstract/10.1103/PhysicsPhysiqueFizika.1.195}{Physics {\bf 1}, 195 (1964).}
		
		\bibitem{belln2} N. Brunner, D. Cavalcanti, S. Pironio, V. Scarani, and S. Wehner, \emph{Bell nonlocality}, \href{https://journals.aps.org/rmp/abstract/10.1103/RevModPhys.86.419}{Rev. Mod. Phys. {\bf 86}, 419 (2014).}
		
		\bibitem{eprs1} H. M. Wiseman, S. J. Jones, and A. C. Doherty, \emph{Steering, Entanglement, Nonlocality, and the Einstein-PodolskyRosen Paradox}, \href{https://journals.aps.org/prl/abstract/10.1103/PhysRevLett.98.140402}{Phys. Rev. Lett. {\bf 98}, 140402 (2007).}
		
		\bibitem{eprs2} S. J. Jones, H. M. Wiseman, and A. C. Doherty, \emph{Entanglement, Einstein-Podolsky-Rosen correlations, Bell nonlocality, and steering}, \href{https://journals.aps.org/pra/abstract/10.1103/PhysRevA.76.052116}{Phys. Rev. A. {\bf 76}, 052116 (2007).}
		
		\bibitem{eprs3} R. Uola, A. C. S. Costa, H. C. Nguyen, and O. Guhne, \emph{Quantum steering}, \href{https://journals.aps.org/rmp/abstract/10.1103/RevModPhys.92.015001}{Rev. Mod. Phys. {\bf 92}, 015001 (2020).}
		
		
		\bibitem{teleport} C. H. Bennett, G. Brassard, C. Crepeau, R. Jozsa, A. Peres, and W. K. Wootters, \emph{Teleporting an unknown quantum state via dual classical and Einstein-Podolsky-Rosen channels}, \href{https://journals.aps.org/prl/abstract/10.1103/PhysRevLett.70.1895}{Phys. Rev. Lett. {\bf 70}, 1895 (1993).}
		
		\bibitem{dense_code} C. H. Bennett, and S. J. Wiesner, \emph{Communication via one- and two-particle operators on Einstein-Podolsky-Rosen states}, \href{https://journals.aps.org/prl/abstract/10.1103/PhysRevLett.69.2881}{Phys. Rev. Lett. \textbf{69}, 2881, (1992).}
		
		\bibitem{qkd} A. K. Ekert, \emph{Quantum cryptography based on Bell’s theorem},
		\href{https://journals.aps.org/prl/abstract/10.1103/PhysRevLett.67.661}{Phys. Rev. Lett. \textbf{67},661, (1991).}  
		
		\bibitem{rand} S. Pironio, A. Acin, S. Massar, A. Boyer de la Giroday, D. N. Matsukevich, P. Maunz, S. Olmschenk, D. Hayes, L. Luo, T. A. Manning, and C. Monroe,\emph{Random numbers certified by Bell’s theorem}, \href{https://www.nature.com/articles/nature09008}{Nature  \textbf{464}, 1021 (2010).} 
		
		
		\bibitem{rac1} M. Pawlowski, and M. Zukowski, \emph{Entanglement-assisted random access codes}, \href{https://journals.aps.org/pra/abstract/10.1103/PhysRevA.81.042326}{Phys. Rev. A {\bf 81}, 042326 (2010).}
		
		\bibitem{prep1}  J. Frohlich, and B. Schubnel, \emph{The preparation of states in quantum mechanics}, \href{https://aip.scitation.org/doi/10.1063/1.4940696}{Journal of Mathematical Physics {\bf 57}, 042101 (2016).}
		
		\bibitem{prep2} D. Girolami, \emph{How Difficult is it to Prepare a Quantum State?}, \href{https://journals.aps.org/prl/abstract/10.1103/PhysRevLett.122.010505}{Phys. Rev. Lett. {\bf 122}, 010505 (2019).}
		
		\bibitem{env1} M. P. Almeida, F. de Melo, M. Hor-Meyll, A. Salles, S. P. Walborn, P. H. Souto Ribeiro, and L. Davidovich, \emph{Environment-Induced Sudden Death of Entanglement}, \href{https://science.sciencemag.org/content/316/5824/579.abstract}{Science {\bf 316}, 579 (2007).}
		
			\bibitem{env4} T. Guha, B. Bhattacharya, D. Das, S. S. Bhattacharya, A. Mukherjee, A. Roy, K. Mukherjee, N. Ganguly, and A. S. Majumdar, \emph{Environmental Effects on Nonlocal Correlations}, \href{http://quanta.ws/ojs/index.php/quanta/article/view/86}{Quanta {\bf 8}, 57 (2019).} 
		
		\bibitem{env2} T. Yu, and J. H. Eberly,  \emph{Sudden Death of Entanglement}, \href{https://science.sciencemag.org/content/323/5914/598}{Science {\bf 323}, 598 (2009).}
		
		
		\bibitem{env3} J.-S. Xu, X.-Y. Xu, C.-F. Li, C.-J. Zhang, X.-B. Zou, and G.-C. Guo, \emph{Experimental investigation of classical and quantum correlations under decoherence}, \href{https://www.nature.com/articles/ncomms1005}{Nat. Commun. {\bf 1}, 7 (2010).}
		
	
		
		\bibitem{env5} K. Modi, A. Brodutch, H. Cable, T. Paterek, and V. Vedral, \emph{The classical-quantum boundary for correlations: Discord and related measures}, \href{https://doi.org/10.1103/RevModPhys.84.1655}{Rev. Mod. Phys. \textbf{84}, 1655 (2012).}
		
		
		
		\bibitem{silva_prl} R. Silva, N. Gisin, Y. Guryanova, and S. Popescu, \emph{Multiple Observers Can Share the Nonlocality of Half of an Entangled Pair by Using Optimal Weak Measurements}, \href{https://doi.org/10.1103/PhysRevLett.114.250401}{Phys. Rev. Lett. \textbf{114}, 250401 (2015).}
		
		\bibitem{CHSH} J. F. Clauser, M. A. Horne, A. Shimony, and R. A. Holt, \emph{Proposed Experiment to Test Local Hidden-Variable Theories}, \href{https://journals.aps.org/prl/abstract/10.1103/PhysRevLett.23.880}{Phys. Rev. Lett. {\bf 23}, 880 (1969).}			
		
		\bibitem{mathematics_mal} S. Mal, A. S. Majumdar, and D. Home, \emph{Sharing of Nonlocality of a Single Member of an Entangled Pair of Qubits Is Not Possible by More than Two Unbiased Observers on the Other Wing},
		\href{https://doi.org/10.3390/math4030048}{Mathematics \textbf{4}, 48 (2016).}
		
		\bibitem{um1} P. Busch, P. Lahti, and P. Mittelstaedt, \emph{The Quantum Theory of
			Measurement}, 2nd ed. (Springer, Berlin, 1996).
		
		\bibitem{unsharp_measurement} P. Busch, M. Grabowski, and P. J. Lathi, \emph{Operational Quantum Physics} (Springer, Berlin, 1997).
		
		\bibitem{igd1} C. A. Fuchs and A. Peres, \emph{Quantum-state disturbance versus information gain: Uncertainty relations for quantum information},\href{https://journals.aps.org/pra/abstract/10.1103/PhysRevA.53.2038}{Phys. Rev. A {\bf 53}, 2038 (1996).}
		
		\bibitem{igd2} F. Buscemi and M. Horodecki, \emph{Towards a Unified Approach to Information-Disturbance Tradeoffs in Quantum Measurements}, \href{https://www.worldscientific.com/doi/abs/10.1142/S1230161209000037} {Open Syst. Inf. Dyn. 16, 29 (2009).}
		
		\bibitem{sasmal_steering} S. Sasmal, D. Das, S. Mal, and A. S. Majumdar, \emph{Steering a single system sequentially by multiple observers},  \href{https://journals.aps.org/pra/abstract/10.1103/PhysRevA.98.012305}{Phys. Rev. A \textbf{98}, 012305 (2018).}
		
		\bibitem{shenoy} A. Shenoy H., S. Designolle, F. Hirsch, R. Silva, N. Gisin, and N. Brunner,  \emph{Unbounded sequence of observers exhibiting Einstein-Podolsky-Rosen steering}, \href{https://journals.aps.org/pra/abstract/10.1103/PhysRevA.99.022317}{Phys. Rev. A {\bf 99}, 022317 (2019).}
		
		\bibitem{choi_steering} Y.-H. Choi, S. Hong, T. Pramanik, H.-T. Lim, Y.-S. Kim, H. Jung, S.-W. Han, S. Moon, and Y.-W. Cho, \emph{Demonstration of simultaneous quantum steering by multiple observers via sequential weak measurements},  \href{https://www.osapublishing.org/optica/fulltext.cfm?uri=optica-7-6-675&id=432421}{Optica \textbf{7}, 675 (2020)}. 
		
		\bibitem{shashank_steering} S. Gupta, A. G. Maity, D. Das, A. Roy, and A. S. Majumdar, \emph{Genuine Einstein-Podolsky-Rosen steering of three-qubit states by multiple sequential observers}, \href{https://journals.aps.org/pra/abstract/10.1103/PhysRevA.103.022421}{Phys. Rev. A {\bf 103}, 022421 (2021).}
		
		\bibitem{steernew21} D. Yao, and C. Ren, \emph{Steering sharing for a two-qubit system via weak measurements}, \href{https://journals.aps.org/pra/abstract/10.1103/PhysRevA.103.052207}{Phys. Rev. A {\bf 103}, 052207 (2021).}
		
		\bibitem{estwo} J. Zhu, M.-J. Hu, G.-C. Guo, C.-F. Li, and Y.-S. Zhang, \emph{Einstein-Podolsky-Rosen steering in two-sided sequential measurements with one entangled pair}, \href{https://doi.org/10.1103/PhysRevA.105.032211}{Phys. Rev. A {\bf 105}, 032211 (2022).}
		
		\bibitem{estwo2} X. Han, Y. Xiao,  H. Qu, R. He, X. Fan, T. Qian, and  Y.  Gu, \emph{Sharing quantum steering among multiple Alices and Bobs via a two-qubit Werner state}, \href{https://doi.org/10.1007/s11128-021-03211-z}{Quantum Inf Process {\bf 20}, 278 (2021).}
		
		\bibitem{npj_2018} M.-J. Hu, Z.-Y. Zhou, X.-M. Hu, C.-F. Li, G.-C. Guo, and Y.-S. Zhang, \emph{Observation of non-locality sharing among three observers with one entangled pair via optimal weak measurement}, \href{https://www.nature.com/articles/s41534-018-0115-x}{npj Quantum Information \textbf{4}, 63 (2018).}
		
		\bibitem{Qu2017} M. Schiavon, L. Calderaro, M. Pittaluga, G. Vallone, and P. Villoresi, \emph{Three-observer Bell inequality violation on a two-qubit
			entangled state}, \href{http://iopscience.iop.org/article/10.1088/2058-9565/aa62be/meta}{Quantum Sci. Technol. \textbf{2} 015010 (2017).} 
		
		
		\bibitem{debarshi_facets} D. Das, A. Ghosal, S. Sasmal, S. Mal, and A. S. Majumdar, \emph{Facets of bipartite nonlocality sharing by multiple observers via sequential measurements}, \href{https://journals.aps.org/pra/abstract/10.1103/PhysRevA.99.022305}{Phys. Rev. A {\bf 99}, 022305 (2019).}
		
		\bibitem{rennew} C. Ren, T. Feng, D. Yao, H. Shi, J. Chen, and X. Zhou, \emph{Passive and active nonlocality sharing for a two-qubit system via weak measurements}, \href{https://journals.aps.org/pra/abstract/10.1103/PhysRevA.100.052121}{Phys. Rev. A {\bf 100}, 052121 (2019).}
		
		\bibitem{Saha19} S. Saha, D. Das, S. Sasmal, D. Sarkar, K. Mukherjee, A. Roy, and S. S. Bhattacharya, \emph{Sharing of tripartite nonlocality by multiple observers measuring sequentially at one side},  \href{https://doi.org/10.1007/s11128-018-2161-x}{Quantum Inf Process {\bf 18}, 42 (2019).}
		
		\bibitem{foletto_entanglement} G. Foletto, L. Calderaro, A. Tavakoli, M. Schiavon, F. Picciariello, A. Cabello, P. Villoresi, and G. Vallone, \emph{Experimental Certification of Sustained Entanglement and Nonlocality after Sequential Measurements},  \href{https://journals.aps.org/prapplied/abstract/10.1103/PhysRevApplied.13.044008}{Phys. Rev. Applied \textbf{13}, 044008 (2020).}
		
		\bibitem{expnew} T. Feng, C. Ren, Y. Tian, M. Luo, H. Shi, J. Chen, and X. Zhou, \emph{Observation of nonlocality sharing via not-so-weak measurements}, \href{https://journals.aps.org/pra/abstract/10.1103/PhysRevA.102.032220}{Phys. Rev. A {\bf 102}, 032220 (2020).} 
		
		\bibitem{zhang21} T. Zhang, and S.-M. Fei, \emph{Sharing quantum nonlocality and genuine nonlocality with independent observables}, \href{https://journals.aps.org/pra/abstract/10.1103/PhysRevA.103.032216}{Phys. Rev. A {\bf 103}, 032216  (2021).}
		
		\bibitem{cglmp} S. Roy, A. Kumari, S. Mal, and A. Sen De, \emph{Robustness of Higher Dimensional Nonlocality against dual noise and sequential measurements}, \href{https://arxiv.org/abs/2012.12200} {arXiv:2012.12200 [quant-ph]}.
		
		\bibitem{prl_brown} P. J. Brown, and R. Colbeck, \emph{Arbitrarily Many Independent Observers Can Share the Nonlocality of a Single Maximally Entangled Qubit Pair}, \href{https://journals.aps.org/prl/abstract/10.1103/PhysRevLett.125.090401}{Phys. Rev. Lett. \textbf{125}, 090401 (2020).} 
		
		
				\bibitem{hallnew} S. Cheng, L. Liu, T. J. Baker, and M. J. W. Hall, \emph{Recycling qubits for the generation of Bell nonlocality between independent sequential observers}, \href{https://doi.org/10.1103/PhysRevA.105.022411}{Phys. Rev. A {\bf 105}, 022411 (2022).}
		
		
		\bibitem{bera_entanglement} A. Bera, S. Mal, A. Sen De, and U. Sen, \emph{Witnessing bipartite entanglement sequentially by multiple observers}, \href{https://journals.aps.org/pra/abstract/10.1103/PhysRevA.98.062304}{Phys. Rev. A
			\textbf{98}, 062304 (2018).}
		
		\bibitem{ananda_entanglement} A. G. Maity, D. Das, A. Ghosal, A. Roy, and A. S. Majumdar, \emph{Detection of genuine tripartite entanglement by multiple sequential observers}, \href{https://journals.aps.org/pra/abstract/10.1103/PhysRevA.101.042340}{ Phys. Rev. A {\bf 101}, 042340 (2020).}
		
		\bibitem{srivastava_entanglement} C. Srivastava, S. Mal, A. Sen De, and U. Sen, \emph{Sequential measurement-device-independent entanglement detection by multiple observers}, \href{https://journals.aps.org/pra/abstract/10.1103/PhysRevA.103.032408}{Phys. Rev. A {\bf 103}, 032408  (2021).}		
		
		\bibitem{coherence_sounak} S. Datta, and A. S. Majumdar, \emph{Sharing of nonlocal advantage of quantum coherence by sequential observers},\href{https://journals.aps.org/pra/abstract/10.1103/PhysRevA.98.042311}{ Phys. Rev. A
			\textbf{98}, 042311 (2018).}
		
		\bibitem{akpan} A. Kumari, and A. K. Pan, \emph{Sharing nonlocality and nontrivial preparation contextuality using the same family of Bell expressions}, \href{https://journals.aps.org/pra/abstract/10.1103/PhysRevA.100.062130}{Phys. Rev. A {\bf 100}, 062130 (2019).}
		
		
		
		\bibitem{rantt} F. J. Curchod, M. Johansson, R. Augusiak, M. J. Hoban, P.  Wittek, and A. Acin, \emph{Unbounded randomness certification using sequences of measurements}  \href{https://journals.aps.org/pra/abstract/10.1103/PhysRevA.95.020102}{Phys. Rev. A {\bf 95}, 020102(R) (2017).}
		
		
		\bibitem{appln1tt} H.-W. Li, Y.-S. Zhang, X.-B. An, Z.-F. Han, and G.-C. Guo, \emph{Three-observer classical dimension witness violation with weak measurement}, \href{https://www.nature.com/articles/s42005-018-0011-x}{Commun. Phys. {\bf 1}, 10 (2018).}
		
		
	
		\bibitem{ractt} K. Mohan, A. Tavakoli, and N. Brunner, \emph{Sequential random access codes and self-testing of quantum measurement instruments}, \href{https://doi.org/10.1088/1367-2630/ab3773}{New J. Phys. 21 083034 (2019).}
		
		
		\bibitem{appln4tt} H. Anwer, S. Muhammad, W. Cherifi, N. Miklin, A. Tavakoli, and M. Bourennane, \emph{Experimental Characterization of Unsharp Qubit Observables and Sequential Measurement Incompatibility via Quantum Random Access Codes}, \href{https://journals.aps.org/prl/abstract/10.1103/PhysRevLett.125.080403}{Phys. Rev. Lett. {\bf 125}, 080403 (2020).}
		
	
		
		\bibitem{appln5tt} G. Foletto, L. Calderaro, G. Vallone, and P. Villoresi, \emph{Experimental demonstration of sequential quantum random access codes}, \href{https://journals.aps.org/prresearch/abstract/10.1103/PhysRevResearch.2.033205}{Phys. Rev. Research {\bf 2}, 033205 (2020).}
				
		\bibitem{sroytt} S. Roy, A. Bera, S. Mal, A. Sen De, and U. Sen, \emph{Recycling the resource: Sequential usage of shared state in quantum teleportation with weak measurements},	\href{https://www.sciencedirect.com/science/article/abs/pii/S0375960121000074}{Phys. Lett. A {\bf 392}, 127143 (2021).}
		
		\bibitem{rsp_shounak} S. Datta, S. Mal, A. K. Pati, A. S, Majumdar, \emph{Remote
			state preparation by multiple observers using a single copy of a two-qubit entangled state}, \href{https://arxiv.org/abs/2109.03682}{arXiv:2109.03682 [quant-ph].}
			
			\bibitem{cctt} A. Tavakoli, and A. Cabello, \emph{Quantum predictions for an unmeasured system cannot be simulated with a finite-memory classical system}, \href{https://journals.aps.org/pra/abstract/10.1103/PhysRevA.97.032131}{Phys. Rev. A {\bf 97}, 032131 (2018).}	
	
		
		\bibitem{bntwo}	S. Cheng, L. Liu, T. J. Baker, and M. J. W. Hall, \emph{Limitations on sharing Bell nonlocality between sequential pairs of observers}, \href{https://doi.org/10.1103/PhysRevA.104.L060201}{Phys. Rev. A {\bf 104}, L060201 (2021).}
		
		\bibitem{cabello} A. Cabello, \emph{Bell nonlocality between sequential pairs of observers}, \href{https://arxiv.org/abs/2103.11844} { arXiv:2103.11844 [quant-ph].}
		
		
		
		\bibitem{rac_debarshi} D. Das,  A. Ghosal, A. G. Maity, S. Kanjilal, and A. Roy, \emph{Ability of unbounded pairs of observers to achieve quantum advantage in random access codes with a single pair of qubits}, \href{https://doi.org/10.1103/PhysRevA.104.L060602}{Phys. Rev. A {\bf 104}, L060602 (2021).}
		
		
		\bibitem{ewo} O. G\"{u}hne, P. Hyllus, D. Bruß, A. Ekert, M. Lewenstein, C. Macchiavello, and A. Sanpera, \emph{Experimental detection of entanglement via witness operators and local measurements}, \href{https://doi.org/10.1080/09500340308234554}{Journal of Modern Optics, \textbf{50},1079 (2003).}
		
		
		
		\bibitem{oew1} O. G\"{u}hne and G. Toth, \emph{Entanglement detection},  \href{https://doi.org/10.1016/j.physrep.2009.02.004}{Phys. Rep. {\bf 474}, 1 (2009).}
		
		\bibitem{rom_prl}	P. Skrzypczyk, and N. Linden, \emph{Robustness of Measurement, Discrimination Games, and Accessible Information,} \href{https://journals.aps.org/prl/abstract/10.1103/PhysRevLett.122.140403}{Phys. Rev. Lett.  \textbf{122},140403 (2019).}
		
		
		
		
		\bibitem{heisenberg} W. Heisenberg, \emph{The Physical Principles of the Quantum Theory}, Chicago Univ. Press (1930).
		
		\bibitem{uno2} P. Busch, T. Heinosaari, J. Schultz, and N. Stevens, \emph{Comparing the degrees of incompatibility inherent in probabilistic physical theories}, \href{https://iopscience.iop.org/article/10.1209/0295-5075/103/10002/meta}{EPL {\bf 103}, 10002 (2013).}
		
		\bibitem{hayashi_book} M. Hayashi, S. Ishizaka, A. Kawachi, G. Kimura, and T. Ogawa, \emph{Introduction to Quantum Information Science}, Springer-Verlag  2015.
	
		
		\bibitem{mlew1} M. Lewenstein, B. Kraus, J. I. Cirac, and P. Horodecki, \emph{Optimization of entanglement witnesses}, \href{https://journals.aps.org/pra/abstract/10.1103/PhysRevA.62.052310}{Phys. Rev. A \textbf{62}, 052310 (2000).}
		
		\bibitem{mlew2} M. Lewenstein, B. Kraus, P. Horodecki, and J. I. Cirac, \emph{Characterization of separable states and entanglement witnesses}, \href{https://journals.aps.org/pra/abstract/10.1103/PhysRevA.63.044304}{Phys. Rev. A \textbf{63}, 044304 (2001).}
		
		\bibitem{prl_nirman} N. Ganguly, S. Adhikari, A.S. Majumdar, and J. Chatterjee, \emph{Entanglement Witness Operator for Quantum Teleportation}, 
		\href{https://journals.aps.org/prl/abstract/10.1103/PhysRevLett.107.270501}{Phys. Rev. Lett. \textbf{107},270501, (2011).}
		
		\bibitem{mems} W. J. Munro, D. F. V. James, A. G. White, and P. G. Kwiat, \emph{Maximizing the entanglement of two mixed qubits}, 
\href{https://doi.org/10.1103/PhysRevA.64.030302}{Phys. Rev. A \textbf{64}, 030302(R) (2001).}

		\bibitem{ewapp1} D. S. Simon, G. Jaeger, and A. V. Sergienko, \emph{Entangled-coherent-state quantum key distribution with entanglement witnessing}, \href{https://doi.org/10.1103/PhysRevA.89.012315}{Phys. Rev. A \textbf{89}, 012315 (2014).}
		
		\bibitem{ewapp2} D. Amaro, M. Muller, and A. K. Pal, \emph{Estimating localizable entanglement from witnesses}, \href{https://doi.org/10.1088/1367-2630/aac485}{New J. Phys. 20 063017 (2018).}
		
	\end{thebibliography}
\end{document}